\documentclass[a4paper,11pt]{article}

\usepackage[utf8]{inputenc}

\usepackage{jheppub}
\usepackage{amsmath, amssymb, amsfonts, mathrsfs, extarrows, mathtools}
\usepackage{graphicx}

\usepackage{tikz}
\usetikzlibrary{arrows,shapes,decorations}
\tikzstyle{block}=[draw opacity=0.7,line width=1.4cm]
\tikzstyle{ell} = [draw,fill=NavyBlue!10,thick,ellipse]
\usetikzlibrary{calc}
\usepackage{calc}
\usepackage{pgfplots}
\usepackage[all]{xy}

\usepackage{tikz-cd}
\usepackage{dsfont}
\usepackage{hyperref}
\usepackage[all]{xy}

\usepackage{tikz-cd}
\usepackage{dsfont}

\usepackage{fixltx2e}

\usepackage{amssymb,amsmath,amsfonts,amsthm,epsfig,pstricks,pst-coil,graphics,pst-node,pst-plot,psfrag}

\newcommand{\be}{\begin{equation}}
\newcommand{\ed}{\end{equation}}

\def\R{\mathbb{R}}

\newcommand{\id}{{\rm id}}

\newcommand{\CFTM}{M}

\newcommand{\ishiket}[1]{|#1\rangle\!\rangle}

\newcommand{\braneket}[1]{\|#1\rangle\!\rangle}

\title{Topological defects and SUSY RG flow}

\author{I. Brunner,}
\author{I. Mayer}
\author{and C. Schmidt-Colinet}
\affiliation{Arnold Sommerfeld Center, Ludwig-Maximilians-Universität\\Theresienstraße 37, 80333 München, Germany}

\emailAdd{Ilka.Brunner@physik.uni-muenchen.de}
\emailAdd{Ingrid.Mayer@physik.uni-muenchen.de}
\emailAdd{cornelius.sc@gmail.com}

\abstract{We study the effect of bulk perturbations of N=(2,2) superconformal minimal models on topological defects. In particular, symmetries and more general topological defects which survive the flow to the IR are identified. Our method is to consider the topological subsector and make use of the Landau-Ginzburg formulation to describe RG flows and topological defects in terms of matrix factorizations.
}

\preprint{LMU-ASC 24/20}

\begin{document}
\maketitle
\flushbottom

\section{Introduction}\label{sec:Introduction}

In this paper we investigate the behavior of topological defects in two-dimensional $N=(2,2)$ supersymmetric conformal field theories under bulk perturbations. The main question we would like to address is which topological defects remain topological under relevant perturbations of the bulk theory. Furthermore, we would like to advertize our method, which is based on flow defects implementing the bulk perturbation in connection with the topological subsector of the model.

Topological defects are natural generalizations of symmetries in conformal field theories and have been studied \textit{e.g.} in \cite{Frohlich:2006ch}.
Topological means that these defects preserve the full $N=(2,2)$ superconformal algebra and can hence be deformed arbitrarily without changing correlation functions -- as long as no operator insertions or other defect lines are hit. Recall that ordinary global symmetries can be reformulated in terms of defects by imposing gluing conditions along a one-dimensional line, relating all fields to their symmetry transformed images. Merging a symmetry defect with its inverse yields the neutral element, given by the trivial line, along which any field is glued to itself. The symmetry transformation of an operator can be implemented by wrapping the symmetry defect line around the operator insertion. Twist fields are modeled as defect changing operators that connect the trivial defect line to the symmetry defect line. In this way, the end of the symmetry defect line automatically has the monodromy properties required of a twist field insertion. The transformation of a boundary condition under the group action is obtained by merging the defect line with the boundary.

The class of topological defects is however in general much bigger than the class of symmetry defects and provides a natural extension of this concept. Arbitrary topological defects still act naturally on local operators as well as boundary conditions, as they can be merged with other objects, as well as with each other, without causing singularities -- as opposed to generic, non-topological defects. Unlike symmetry defects, general topological defects will not have an inverse. However, there is a defect of opposite orientation and fusion of the initial defect with the orientation reversed version contains the identity. 


We would like to investigate symmetries and their generalizations under bulk perturbations. In general, a symmetry can be incompatible with a given bulk perturbation, and the perturbation breaks it. A bulk perturbation forces a defect or boundary to re-adjust; in the case of defects, the reflectivity and transmissivity will change along the flow induced from the bulk. We are interested in specifying preserved symmetry defects and more general topological defects that remain topological under perturbations.

We would like to address this question using a non-perturbative method, namely, RG interfaces. Here, the basic idea is to connect the initial (UV) theory and the perturbed (IR) theory by a defect line. To construct these defects, one restricts the perturbation to a finite region of the two-dimensional surface on which the theory is formulated. The perturbation and the resulting RG flow then drives the theory inside the restricted domain to the new IR theory, whereas the theory stays in the UV in the other parts of the surface. In this way, a one-dimensional domain wall between the two regions is created, to which we refer as RG defect. 

Having constructed the RG defect, one can apply it to investigate for example the behavior of boundary conditions under bulk perturbations. For this, one merges the RG defect with some initial UV boundary condition and obtains a new boundary condition in the IR. This procedure amounts to first limiting the perturbation to a region away from the boundary, then extending it onto the boundary. In practice, the merging is highly singular on the level of the full conformal field theory, see \cite{Bachas:2007td,Konechny:2015qla} for results on the case of free fields and marginal deformations.

To circumvent these problems, we want to consider the topological subsector of $N=(2,2)$ superconformal field theories. On the level of the topological theory, the merging of defects is well-defined and naturally smooth. Still, using that the topological model describes a subsector of the original theory protected by supersymmetry, one can draw conclusions about the full superconformal theory. 

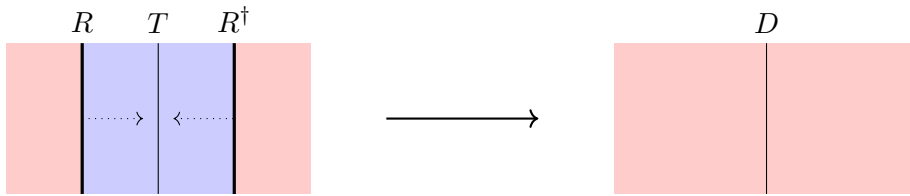
\begin{figure}[h]\label{Intro_BasicPicture}
\begin{center}
\begin{tikzpicture}
\fill[fill=red!20!white] (0,0) -- (0,2) -- (1,2) -- (1,0) -- cycle; 
\fill[fill=blue!20!white] (1,0) -- (1,2) -- (3,2) -- (3,0) -- cycle;
\fill[fill=red!20!white] (3,0) -- (3,2) -- (4,2) -- (4,0) -- cycle; 
\draw[very thick](1,0) -- (1,2) node[above]{$R$}; 
\draw (2,0) -- (2,2) node[above]{$T$}; 
\draw[very thick](3,0) -- (3,2) node[above]{$R^\dagger$}; 
\draw[dotted, ->] (1,1) -- (1.8,1); 
\draw[dotted, <-] (2.2,1) -- (3,1); 
\draw[thick, ->] (5,1) -- (7,1); 
\fill[fill=red!20!white] (8,0) -- (8,2) -- (12,2) -- (12,0) -- cycle; 
\draw (10,0) -- (10,2) node[above]{$D$}; 
\end{tikzpicture}
\end{center}
\caption{Flow of a UV topological defect $T$ under a bulk perturbation, for a case where this leads to an IR defect $D$. The flow is modeled as a fusion product of an RG defect $R$ and its conjugate $R^\dagger$ with $T$.}
\end{figure}

For the case of topological theories, \cite{Klos:2019axh} describe how the IR theory can be embedded into the UV theory by an orbifold-like construction \cite{Frohlich:2006ch, Carqueville:2012dk}, thereby describing many important properties of RG defects and their fusion. For the question we have in mind, it is important that RG defects allow us to keep a perturbation away from some UV data, whose behavior under the flow then amounts to merging and contracting RG defects. To be concrete, we enclose a defect line representing a symmetry defect of the UV theory between RG defect lines and then merge all lines to get a resulting IR defect (see Figure~\ref{Intro_BasicPicture}).

As an example, we consider the case of $N=(2,2)$ minimal models, whose topological subsector can be described by an orbifold of a Landau-Ginzburg model with superpotential $W=X^d$. The defects in this theory are completely under control, as they can be formulated in terms of matrix factorizations in orbifold theories. It has also been elaborated on how to take the fusion product \cite{Brunner:2007qu, Brunner:2007ur} and the RG defects were constructed in \cite{Brunner:2007ur} and further investigated in \cite{Klos:2019axh}. Furthermore, the symmetries of the conformal field theory can be lifted to the topological sector, and their description in terms of defects is also well-known \cite{Brunner:2007ur}. This immediately allows to investigate symmetries from the point of view of RG defects.

To generalize to arbitrary topological defects, where topological now refers to a property of the full conformal field theory, we make use of an interplay between conformal field theory and matrix factorization techniques, identifying CFT topological defects with specific matrix factorizations. This is possible because the theory is a minimal model. Conclusions about generically preserved topological interfaces can then be obtained by fusing on the level of the topological model.

This paper is organized as follows. In section \ref{sec:DefRG} we summarize properties of RG defects as well as different properties that other, in particular topological defects, might have with respect to them. As explained above, one question to answer is the behavior of a given defect under a bulk induced flow and to isolate defects that remain topological. A different and stronger property of a defect can be that the perturbation is totally ``invisible" to the topological defect, and hence the defect commutes with the RG defect. Such defects were considered in \cite{Gaiotto:2012np}, where they found an application  for the construction of RG defects. As follows easily from their defining properties, this class of defects closes under fusion and their algebraic relations are the same in UV and IR.

In section \ref{sec:A-picture}, we introduce our example, namely $N=(2,2)$ minimal models, and briefly review the bulk perturbations and behavior of D-branes under bulk perturbations in the mirror A-picture, following \cite{Hori:2000ck_HIV}. In section \ref{sec:MF} we review the description of defects in terms of matrix factorizations on orbifolds in the B-picture. In particular, we introduce the matrix factorizations corresponding to topological defects of the full conformal field theory and we review the RG defects of \cite{Brunner:2007ur, Klos:2019axh}. The basic data contained in an RG defect of a topological theory is which sectors decouple along the flow, in particular, which vacua or boundary states. Finally, in section \ref{sec:Results} we apply matrix factorization techniques to obtain results on symmetries and general topological defects in $N=2$ minimal models. More concretely, we identify symmetry defects that commute with RG defects, and we show that no other topological defect can \emph{commute} with the flow defects. Our argument makes heavily use of the fact that some of the supersymmetric vacua decouple from the superconformal theory along the flow and become massive. However, there are more defects that stay topological under a bulk induced flow, and we identify a symmetry defect (other than the one implementing supersymmetry) that remains invariant under any perturbation in any minimal model. As we show, as a consequence, there is one further defect that is not a symmetry defect, yet is also preserved under any flow. This defect can be interpreted as a bound state of the identity defect with the preserved symmetry defect, or in other words, it is obtained using the cone construction from two symmetry defects.

Finally, we interpret the results back in the initial CFT in \ref{sec:BackCFT}. We will in particular see that the generically preserved defects amount to spectral flow symmetries. In section \ref{sec:conclusion} we summarize and conclude.

\section{Defects and RG flows}\label{sec:DefRG}

Consider an initial (UV) conformal field theory with a topological defect $T$. A natural question is whether the defect remains topological under perturbations of the bulk theory. For a discussion of how topological defect lines can constrain RG flows, see \cite{Chang:2018iay}. The topological defects in the UV theory come with a natural product structure obtained by merging of defect lines. One might therefore also be interested to ask whether such algebraic structures are preserved for suitable subclasses of defects and deformations. 

On the level of the full conformal field theory, topological defects are distinguished by the property that the reflectivity is $0$, while the transmissivity is $1$, see \cite{Quella:2006de} for definitions. In principle, one could try to investigate these conditions under bulk perturbations using conformal perturbation theory, see \cite{Makabe:2017uch} for some results.

Alternatively, the RG flow can be described using RG defects \cite{Brunner:2007ur,Klos:2019axh}. Here, the basic idea is to initially restrict the perturbation to a domain which is a proper subset of the two-dimensional spacetime on which the theory is formulated. The RG flow on this perturbed domain drives the theory to the IR, whereas it stays in the UV outside of it. In this way, a domain wall separating UV and IR is created along the boundary of the perturbed domain. 

This procedure can  be applied to determine for instance the RG images of UV boundaries. For this, one  "merges" the RG interface with the UV boundary conditions, obtaining valid boundary conditions of the IR theory. One advantage of this way of regularizing combined bulk-boundary flows  is that it is per construction non-perturbative in the bulk coupling constants. An obvious  disadvantage is that in general it is difficult to construct the defects explicitly, and even if they are known, the procedure of "merging" is highly singular and difficult to control. For explicit constructions of flow defects in conformal field theory see \cite{Gaiotto:2012np,Konechny:2016eek,Stanishkov:2016pvi,Poghosyan:2014jia}.

$N=(2,2)$ superconformal theories provide a class of theories where the outlined procedure can be implemented in practice \cite{Brunner:2007ur,Becker:2016umn} . Here, one can pick out a topological sector and define the flow defects on the topological level.  This is possible if the perturbation preserves one half of the supersymmetry and  the topological sector can be chosen accordingly. The flow defects can thus be studied together with other defects or boundaries compatible with the same supersymmetry. The merging of defects is a well-defined procedure on this level, where we  restrict to a topological subsector that allows a non-singular fusion product. Since this protected sector is  a subsector of the full superconformal theory, one can draw conclusions also for the full theory. 

A defining property of an RG defect $R$  is that the fusion of the defect with its orientation reversed defect $R^\dagger$ yields the invisible, trivial  defect $\id_{IR}$ of the IR theory \cite{Klos:2019axh}:
\begin{equation}\label{eq:RdaggerR}
R \ast R^\dagger = \id_{IR} \ .
\end{equation}
On the other hand, the fusion in opposite order results in a projection defect of the UV theory,
\begin{equation}
R^\dagger \ast R =P \; ,
\end{equation}
where the identity $P\ast P=P$ follows directly from (\ref{eq:RdaggerR}). Enclosing a strip of IR theory in between the UV theory, the fusion product will in general result in a projector (not equal to the identity), as some of the information of the UV is "forgotten" in the squeezed-in region. To determine the IR image of a UV defect $D_{UV}$, one has to take the fusion product
\begin{equation}
D_{IR} = R\ast D_{UV} \ast R^\dagger \ .
\end{equation}
We are particularly interested in the case that $D_{UV}$ is a topological or, more strongly, a symmetry defect on the level of the \emph{full} conformal field theory, and we are interested in finding cases were the resulting defect is again topological. To do so, we need to establish the class of (full CFT) topological defects inside the class of defects of the topological theory (which  naturally are \emph{all} topological), thereby carrying over information from the full conformal field theory. We use the topological  theory merely as a tool for computing the fusion products.

Note that in the above computation of the IR image of a UV defect the product $R\ast T$ ($T$ a topological defect) would be well-defined and non-singular even on the level of the full conformal theory, as $T$ could be moved on top of the flow defect without singularities. But the final step, fusing $(RT)$ with $R^\dagger$ would  be highly singular in general. 

Our interest is in pairs $(T_{IR}, T_{UV})$ of topological defects related by RG flow, thus satisfying
\begin{equation}\label{eq:flowtop}
T_{IR}= R \ast T_{UV} \ast R^\dagger \ .
\end{equation}
Fusing this identity with $R$, one obtains
\begin{equation}
T_{IR} \ast R = R \ast T_{UV} \ast P \; .
\end{equation}
Starting from here, it is natural to impose the condition that the topological $UV$ defect commutes with the projection operator:
\begin{equation}
P\ast T = T \ast P \ .
\end{equation}
Combining this with the condition (\ref{eq:flowtop}), we obtain an intertwining property for this special class of defects:
\begin{equation} \label{eq:intertwine}
T_{IR} \ast R = R \ast T_{UV} \ ,
\end{equation}
which is stronger than (\ref{eq:flowtop}).
All fusion products in this equation can be taken in a smooth manner on the level of the full CFT, and hence this equation can be directly translated to CFT. It means that the RG defect is completely transparent for the topological defect.  The topological defect  can be "pulled through" the RG defect, changing only from $T_{UV}$ to $T_{IR}$, and it can also intersect with $R$ in a topological manner, even on the level of the full conformal field theory. Such defects were used in \cite{Gaiotto:2012np} for the actual construction of the RG defects.

We would like to emphasize that the set of topological defects satisfying (\ref{eq:intertwine}) closes under fusion, see also the analysis in \cite{Chang:2018iay}. Indeed, the full algebraic structure is mapped from UV to IR. This is different for the set of defects satisfying only (\ref{eq:flowtop}) which does not close under fusion. However, the intertwining defects (\ref{eq:intertwine}) operate naturally on the defects (\ref{eq:flowtop}), such that the defects (\ref{eq:flowtop}) can be organized in orbits of (\ref{eq:intertwine}).

In the following chapter, we will work out an example for the structures described above, namely supersymmetric minimal models. The topological sector of them is described by Landau-Ginzburg models with superpotential $W=X^d$. Flows to Landau-Ginzburg models with superpotential $W=X^{d'}$, $d'<d$, are described by  interfaces preserving A-type supersymmetry. We later take a mirror perspective and consider Landau-Ginzburg orbifolds, where the orbifold group is   ${\mathbb Z}_d$.  Defects in the  B-type models are then described by suitable equivariant matrix factorizations. The RG defects have been determined in \cite{Brunner:2007ur} and it is also known how to describe the CFT-topological defects in terms of LG defects. Thus, the program outlined above can be completed in this class of examples.

To summarize our results, we show that certain symmetry defects satisfy (\ref{eq:intertwine}) and that there are no other (non-symmetry) topological defects that do so. We also exhibit two symmetry defects and one other topological defect satisfying the weaker condition (\ref{eq:flowtop}) \emph{for all flows}. In models with intertwining symmetry defects, one can generate more defects satisfying (\ref{eq:flowtop}) by acting on the generic ones. In this way, we organize  (\ref{eq:flowtop})  into orbits under the preserved symmetry.





\section{Review of minimal models, RG flows and branes in the A-picture}\label{sec:A-picture}


For our discussion of topological defects, we focus on $N = (2,2)$ superconformal minimal models, for which a Landau-Ginzburg formulation exists \cite{Vafa:1988uu}. A Landau-Ginzburg model in a finite number of chiral superfields $X_i$ is described by an action of the form
\begin{equation}
S = S_D + S_F \; ,
\end{equation}
where the D-term is given by
\begin{equation}\label{LGaction}
S_D = \int d^2 x \; d^4 \theta  \; K (X_i, \bar{X}_i) \; ,
\end{equation}
and we consider a flat Kähler potential of the form $K = \sum_i X_i \bar{X}_i$. The F-term is given by
\begin{equation}
S_F = \int d^2 x \; d\theta^- d\theta^+ \; W(X_i) \big\vert_{\bar{\theta}^\pm = 0} \; + \; \int d^2 x \; d \bar{\theta}^+ d \bar{\theta}^- \; \bar{W} (\bar{X}_i) \big\vert_{\theta^\pm = 0} \; ,
\end{equation}
with $W$ a holomorphic function in the chiral superfields, referred to as superpotential. In the absence of boundaries or defects, the model described by the action \eqref{LGaction} is invariant under the $N = (2,2)$ supersymmetry transformations
\begin{equation}
\delta = \epsilon_+ Q_- - \epsilon_- Q_+ - \bar{\epsilon}_+ \bar{Q}_- + \bar{\epsilon}_- \bar{Q}_+ \; ,
\end{equation}
where $Q_\pm$ and $\bar{Q}_\pm$ denote the four supercharges. Adding additional objects to the model typically breaks the supersymmetry of the bulk theory. In the presence of boundaries, at most half of the supersymmetries of the bulk theory can be preserved. As opposed to boundaries, special classes of defects can preserve the full supersymmetry of the bulk model. \\
For $N = (2,2)$ Landau-Ginzburg models with quasi-homogeneous superpotential, the IR fixed points of RG flows are uniquely determined by the exact superpotential and can be identified with $N = (2,2)$ superconformal field theories \cite{Vafa:1988uu}. The defects preserving the full supersymmetry flow to topological defects of the superconformal model.
We will consider Landau-Ginzburg models with single chiral superfield $X$ and superpotential $W = X^d$, which correspond to superconformal minimal models $\mathcal{M}_{k}$ at level $k = d-2$ with A-type modular invariant partition function and central charge $c = (3 - 6/d)$ at the IR fixed point. These models admit $d-1$ relevant chiral perturbations, which preserve A-type supersymmetry, and induce an RG flow to some model determined by $W = X^{d'}$, with $d' < d$. A-twisted Landau-Ginzburg models with superpotential $W = X^d$ are related by mirror symmetry to B-twisted $\mathbb{Z}_d$-orbifolds thereof. We denote these models by $X^d / \mathbb{Z}_d$. The corresponding IR fixed point SCFT is the $\mathbb{Z}_{k+2}$-orbifold of $\mathcal{M}_{k}$, which we denote by $\mathcal{M}_{k}/\mathbb{Z}_{k+2}$. The $d-1$ relevant perturbations of the orbifold model are generated by twisted chiral fields and preserve B-type supersymmetry. In the following, we will discuss RG flows, boundary conditions and topological defects in the A-model.

\subsection{RG flow in terms of the superpotential}
Let us first recall an effective description of an RG flow starting from a perturbation of a UV minimal model. Our starting point in the UV is the theory with superpotential $W = X^{d}$. Relevant chiral operators of the corresponding superconformal minimal model, which at the UV fixed point correspond to the monomials $1,\, X,\ldots,\, X^{k=d-2}$, can be employed to perturb the SCFT away from this superconformal fixed point. This leads to a supersymmetric quantum field theory with superpotential
\begin{equation}\label{EffectivePerturbedW}
W(\lambda)=X^d+\sum_{j=1}^{d-2}\lambda_j X^j\,.
\end{equation}
Here, the $\lambda_j$ are functions of the coupling constants of the initial perturbation by the relevant chiral operators, see {\it e.g.}~\cite{Warner:1993zh}. The perturbed superpotential $W(\lambda)$ does generically not have degenerate critical points. However, if a degenerate critical point persists (corresponding to a fine tuning of the initial coupling constants), there exists an RG flow to an IR theory which is again a non-trivial SCFT. Due to non-renormalization theorems of supersymmetric QFT, the RG flow of the topological sector of a perturbed theory follows from a rather simple rescaling argument~\cite{Cecotti-Vafa-ttbar,Vafa:2001ra}: Under a change of scale $z\mapsto z/\xi$, $\theta\mapsto \sqrt{\xi}\theta$ for a dimensionless factor $\xi$,
\begin{equation}
W(\lambda)\mapsto \xi W(\lambda)\,.
\end{equation}
This behavior of the superpotential is tied to the fact that under the same change of scale, the operators of the chiral ring will behave in a way such that the monomials will only undergo a rescaling of the form
\begin{equation}\label{xrescaling}
X^j\mapsto \xi^{j/d}X^j\,,
\end{equation} 
and the coefficients transform as
\begin{equation}
\lambda_j\mapsto \xi^{1-j/d}\lambda_j\,.
\end{equation}
The IR fixed point for the perturbed theory with superpotential~\eqref{EffectivePerturbedW} is obtained in the limit $\xi\rightarrow\infty$, possibly by an appropriate reparametrization of monomials and coefficients (fields and couplings), and a shift in the superpotential. If we mark the location of critical points of $W(\lambda)$ in the complex plane, this reparametrization and shift serves to keep the IR fixed point at the origin of the complex plane, while the locations of the other critical points of $W$ are pushed off to infinity as $\xi\rightarrow\infty$.\footnote{In the following we will sometimes refer to a mark in the plane where derivatives of $W$ vanish as ``a critical point'', without regard to its multiplicity.} In this way one can establish a flow to an IR fixed point corresponding to a minimal model with superpotential $W=X^{d'}$ for $d'<d$.\footnote{In particular, there is always a perturbation for which~\eqref{EffectivePerturbedW} has a critical point where $d-2$ derivatives vanish; by a shift in the field $X$ and in the superpotential, the latter can be cast in the form $W=X^d+\lambda X^{d-1}$, which flows to a minimal model with superpotential $W=X^{d'=d-1}$.}

\subsection{Branes and flows}\label{sec:branesandflows}

\begin{figure}[h]
\centering
 \begin{tikzpicture}
\def\R{2cm}
\draw (0,0) circle[radius=\R];
\draw (0,0) -- (0:\R)    (0,0) -- (45:\R)     (0,0) -- (90:\R)   (0,0) -- (135:\R)    (0,0) -- (180:\R)     (0,0) -- (225:\R)   (0,0) -- (270:\R) (0,0) -- (315:\R);
\draw[red, thick, ->] (0:\R) -- (0:\R/1.7);
\draw[red, thick] (0:\R/1.7) -- (0,0);
\draw[red, thick, ->] (0:0) -- (45:\R/1.4);
\draw[red, thick] (45:\R/1.4) -- (45:\R);
\draw[green, thick, ->] (315:\R) -- (315:\R/1.7);
\draw[green, thick] (315:\R/1.7) -- (0,0);
\draw[green, thick, ->] (0,0) -- (90:\R/1.4);
\draw[green, thick] (90:\R/1.4) -- (90:\R);
\node[font=\footnotesize] at ( 0:\R*1.3) {$b+1$};
\node[font=\footnotesize] at ( 45:\R*1.3) {$b+2$};
\node[font=\footnotesize] at ( 90:\R*1.3) {$b+3$};
\node[font=\footnotesize] at ( 135:\R*1.3) {$b+4$};
\node[font=\footnotesize] at ( 180:\R*1.3) {$b+5$};
\node[font=\footnotesize] at ( 225:\R*1.3) {$b+6$};
\node[font=\footnotesize] at ( 270:\R*1.3) {$b+7$};
\node[font=\footnotesize] at ( 315:\R*1.3) {$b$};
\filldraw (0,0) circle[radius=0.05];
\node[font=\footnotesize] at ( 200:\R/5) {$x_c$};
\end{tikzpicture}
\caption{Pictorial representation for elementary branes for the case $W = X^8$. The red lines represent an elementary brane $(b+1, x_c, b+2)$, the green lines represent a bound state $(b, x_c, b+3)$ of three neighboring branes corresponding to smallest wedges.}\label{ExamplePictorialRepres1}
\end{figure}
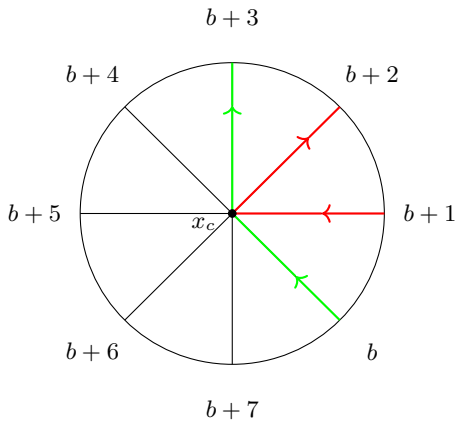

The A-branes in minimal models are described in  Landau-Ginzburg language by specifying one-dimensional submanifolds of ${\mathbb C}$, the target space parametrized by the field $X$. Admissible submanifolds are composed of straight lines, along which the imaginary part of the superpotential remains constant, and the real part of the superpotential is bounded from below \cite{Hori:2000ck_HIV,Govindarajan:2000my}. The lines emanate from the critical points of $W$. At the conformal fixed point, $W=X^d$ and there is only one single critical point at the origin, whereas for a generic degree $d$ superpotential there are $d-1$ critical points. The admissible lines $\gamma$ run from the origin along $z=t e^{2\pi i b/(k+2)}$, $d=k+2$, 
parametrized by $t\in\mathbb{R}_+$, for $b=0,1,\ldots,k+1$. This means that the plane parametrized by $X$ is divided into $d$ segments. Every elementary boundary state
corresponds to a pair of these lines, and following \cite{Brunner:2007ur} we will label them by
$(b_1,x_c,b_2)$, where $x_c$ is the critical point. Indeed, there is a set of smallest segments, corresponding to $(b, x_c, b+1)$.  The branes corresponding to unions of neighboring segments, and thus to other elementary branes, can be thought of as bound states of the branes corresponding to smallest segments, see figure \ref{ExamplePictorialRepres1} for an example. For instance, in \cite{Hori:2000ck_HIV} the Witten index was computed for all such boundary conditions and it was found that it is $\pm 1$ for two neighboring smallest wedges $(b,x_c, b+1)$ and $(b+1, x_c, b+2)$. The bound state $(b, x_c, b+2)$ is formed by perturbing with the corresponding relevant boundary field, flowing to a new boundary condition.  In the graphical representation by wedges, each of the two constituent branes $(b, x_c, b+1)$ and $(b+1, x_c, b+2)$ contains a line along $e^{2\pi i (b+1)/(k+2)}$, but with different orientation. The ``cancellation" of such lines corresponds to the  bound state formation in field theory.

Under deformations of the superpotential by lower order polynomials, the single critical point at the origin splits up. As explained above, at the endpoint of the flow, we obtain a Landau-Ginzburg model with a superpotential of lower order and a critical point at the origin. The other critical points are driven to $\infty$ and the corresponding vacua decouple from the non-trivial IR fixed point. A-branes attached to such a critical point likewise decouple. A-branes attached to a critical point remaining at the origin become A-branes of the new IR fixed point of the perturbed theory. 

To illustrate this, let us consider the simplest case, namely, the flow from $W=X^d$ to $W=X^{d-1}$ described by the perturbation $W=X^{d} + \lambda X^{d-1}$. The UV superpotential at $\lambda=0$ has a critical point of order $d-1$ at the origin, which for $\lambda \neq 0$ splits into a critical point of order  $d-2$  at the origin and a critical point of order $1$ at $X\sim \lambda$. Accordingly, one A-brane corresponding to one specific wedge of minimal angle $2\pi i/d$ or lines $(b_d, 0, b_d+1)$ has to decouple from the theory. The possible flows are specified completely by giving the A-brane that flows off to infinity. There are $d$ possibilities, corresponding precisely to a choice of phase for the perturbation. 
Let us consider a bound state $(b_d, 0, b_d+2)$ of the decoupling brane $(b_d, 0, b_d+1)$ and the brane $(b_d+ 1, 0, b_d+2)$. Along the flow, $(b_d, 0, b_d+1)$ decouples and thus the brane $(b_d, 0, b_d+2)$ flows to a single wedge brane in the IR theory. 

Generically, there are no symmetries preserved along the flow. The symmetry group of the UV is ${\mathbb Z}_d$ and the symmetry group of the IR is ${\mathbb Z}_{d'}$ and to have a preserved symmetry, these two groups need to have a common subgroup. Provided that this is the case, there exist symmetry preserving flows which, from the point of view of the A-branes, are provided by the property that the set of decoupling A-branes is symmetric with respect to the preserved symmetry group, see \cite{Gaberdiel:2007us} for a previous analysis of D-branes along these flows. A discussion of the flows in the B-picture can be found in section \ref{chap:commsymm}.

\subsection{Minimal model RCFT}

The Landau-Ginzburg models with superpotential $W=X^d$ have a description in terms of rational conformal field theory, as we briefly recall. For the diagonal models based on the coset
\begin{equation}\label{eq:CosetN2}
\frac{\mathfrak{su}(2)_k\oplus\mathfrak{u}(1)_2}{\mathfrak{u}(1)_{k+2}}\,,\qquad k \in \mathbb{N}\,,
\end{equation}
the central charge is 
\begin{equation}
c=\frac{3k}{k+2}\,.
\end{equation}
The bosonic submodules of  full superconformal modules at level $k$ are 
labeled by triples of integers $(l,m,s)$ with
\begin{align}\label{N2labels}
&0\le l \le k\,,\quad-k-1\le m\le k+2\,,\quad-1\le s\le 2\,,\quad l+m+s = 0 \text{ mod }2\,.
\end{align}
The representations labeled by $(l,m,s)$ and $(k-l,m+k+2,s+2)$ are identical. Full superconformal representations
in the Ramond sector are isomorphic to the direct sum $(l,m)\equiv (l,m,-1)\oplus(l,m,1)$ (where $l+m$ is odd), and full 
Neveu-Schwarz representations are given by $(l,m)\equiv(l,m,0)\oplus(l,m,2)$ (where $l+m$ is even).
For a bosonic subalgebra representation with labels $(l,m,s)$ in the ``standard range'' $|m-s|\le l$, 
the highest weight is
\begin{equation}\label{N2h}
h_{l,m,s} = \frac{l(l+2)-m^2}{4(k+2)}+\frac{s^2}{8} \,,
\end{equation}
and the $U(1)$ R-current charge of the highest weight state is
\begin{equation}\label{N2q}
q_{l,m,s}=\frac{m}{k+2}-\frac{s}{2}\,.
\end{equation}
If the labels are not in the standard range, \eqref{N2h} holds up to integers, and~\eqref{N2q} up to even integers.

The representation spaces for the $N=2$ supersymmetry algebra are given by
\begin{equation}
{\cal H}_{l,m}= {\cal H}_{l,m,s} \oplus {\cal H}_{l,m,s+2} \; ,
\end{equation}
where $s$ even corresponds to the NS sector and $s$ odd to the R sector. On the level of the full conformal field theory, we must include left and right movers, which we distinguish by a tilde. We consider a supersymmetric theory (as the LG model is supersymmetric) together with a specific action of $(-1)^F$, in our case $(-1)^F=e^{\pi i J_0-\pi i \tilde{J}_0}$. When projecting with the chosen $(-1)^F$, one obtains a CFT model with modular invariant partition function which is 
\begin{equation}
Z=\sum \chi_{l,m,s}(q) \bar\chi_{l,m,s}(\bar{q}) \; .
\end{equation}
Other choices of $(-1)^F$ lead to other modular invariant partition functions, where the spin labels of left and right movers are paired up differently. 

A-type boundary states have to satisfy the following gluing conditions:
\begin{equation}\label{eq:Aglue}
\begin{split}
(L_n-\tilde{L}_{-n})\braneket{A} &=0 \; , \\ 
(J_n-\tilde{J}_{-n})\braneket{A} &=0 \; ,\\
(G^\pm_r+i\eta\, \tilde{G}_{- r}^\mp)\braneket{A} &=0 \; ,
\end{split}
\end{equation}
where the sign $\eta$ implements the spin structure. For the case of the GSO-projected minimal model, we can apply Cardy's formula to find the solutions
\begin{equation}\label{CardyBoundaryState}
\braneket{L,M_{CFT},S}_{A}=(2k+4)^{1/4}\sum_{l=0}^k\sum_{m=-l}^{l+1}\sum_{s}
\frac{S_{Ll}}{\sqrt{S_{0l}}}\,e^{\frac{i\pi mM_{CFT}}{k+2}}e^{-\frac{i\pi Ss}{2}}\ishiket{l,m,s}_{A}\,,
\end{equation}
where in the sum $s$ runs over the labels $-1,0,1,2$ with $l+m+s$ even. The GSO projection can be made undone, essentially considering NS and R sector separately. The
boundary states~\eqref{CardyBoundaryState} satisfy the above boundary conditions with
spin structure $\eta=e^{i\pi S}$. In the formula, the $\mathfrak{su}(2)_k$ modular $S$-matrix has entries
\begin{equation}
S_{Ll}=\sqrt{\frac{2}{k+2}}\,\sin\left(\frac{\pi (L+1)(l+1)}{k+2}\right)\,.
\end{equation}


To match the CFT description with the LG description, note that the model $W=X^d$ corresponds to the CFT model with $k+2=d$. Under the correspondence, the powers of the chiral field $X$ get mapped to chiral primaries, $X^l \leftrightarrow \phi_{(l,l,0)}\tilde\phi_{(l,l,0)}$, where $\phi_{(l,l,0)}$ denotes a field with coset labels $(l,l,0)$ and the tilde refers to the right moving part. Note that the actual perturbing field that induces the perturbation is a $G$-descendant of this field, which then carries representation labels $(l,l,2)$ for left and right movers. 

As has been worked out in \cite{Hori:2000ck_HIV}, the description of the Landau-Ginzburg branes can be mapped to A-type D-branes in conformal field theory. For this, one first chooses a fixed spin structure and fixes the $S$-labels of the boundary states to be odd. 
The dictionary between the
coset labels and the lines is then
\begin{equation}
L+1=|b_2-b_1|\,,\qquad M_{CFT}=b_1+b_2\,,\qquad S\propto \rm{sign} (b_2-b_1)\,.
\end{equation}
Boundary states of fixed $L,M_{CFT}$ with  $S$-labels shifted by two correspond to antibrane pairs. In the Landau-Ginzburg picture, this is mapped to a reversal of orientation of the line describing the boundary condition. Moreover, the $L$-label of the CFT determines the size of the segment, the $L=0$ Cardy boundary states correspond to the elementary wedges of an angle $\frac{2\pi}{k+2}$. The individual smallest wedges are then labeled by $M_{CFT}$. For fixed spin label $S$ and $L=0$, $M_{CFT}$ can take $k+2$ different values, $M_{CFT}=1,3, \dots 2k+3$. The ${\mathbb Z}_{k+2}$-symmetry of the model is generated by a rotation by an angle $\frac{2\pi}{k+2}$ and shifts the $M_{CFT}$-labels. 
Higher $L$ branes can be obtained as bound states of such branes, a basic check is provided by the conservation of RR charges along this boundary flow, see \cite{Gaberdiel:2007us}. 


\subsection{Topological defects and RG flow}

Topological defects preserving the full $N=2$ supersymmetry can be constructed following \cite{Petkova:2000ip}. Note that these defects preserve two copies of the $N=2$ algebra, which contains in particular the stress energy tensor. This means that the defects we discuss in this section are topological in the sense of the full conformal field theory and can for example be merged smoothly and also act on boundary conditions. They intertwine the action of the full superconformal algebra:
\begin{equation}
L_n^{(1)} {\cal D} = {\cal D} L_n^{(2)}, \quad J_n^{(1)} {\cal D} = {\cal D} J_n^{(2)}, \quad G_n^{(1)\pm} {\cal D} = {\cal D} G_n^{(2)\pm},
\end{equation}
and likewise for the right movers.
In the fully supersymmetric (i.e. non-GSO-projected) theory, they are given by 
\begin{equation}
{\cal D}_{LMS\tilde{S}}=\sum_{(l,m)}\sum_{s,\tilde{s}}D_{LMS\tilde{S}}^{lms{\tilde{s}}}\,\|l,m,s,\tilde{s}\| \; ,
\end{equation}
with~\cite{Brunner:2007ur}
\begin{equation}\label{quantumdimensions}
D_{LMS\tilde{S}}^{lms{\tilde{s}}}=e^{-\frac{i\pi}{2}\tilde{S}(s+\tilde{s})}\frac{S_{(L,M,S-\tilde{S})(l,m,s)}}{S_{(0,0,0)(l,m,s)}}\,.
\end{equation}
Here $(L,M,S)$ must be a valid coset label, and $S-\tilde{S}$ is even. Furthermore, $\|l,m,s,\tilde{s}\|$ acts as a projection operator on the Hilbert space with the respective representation labels. 
Identification rules for the defect labels read
\begin{equation}
(L,M,S,\tilde{S})=(k-L,M+k+2,S+2,\tilde{S}+2)\,.
\end{equation}
In the GSO-projected theory, there is only one $S$-label and we have
\begin{equation}
{\cal D}_{LMS}=\sum_{([l,m,s])} D_{LMS}^{lms}\,\|l,m,s\| \; .
\end{equation}
Here, we will fix all $S$-labels to be $0$, indeed, we will discuss $S=2$ separately. We want to regard the defects as maps on the boundary states. For the latter,  we have an LG interpretation and interpreting the defects as functors on the boundary category allows us to directly carry over their description to the LG picture. To match the boundary conditions, we previously fixed $S=1$, and maintaining this means to omit odd $S$ for the discussion of defects. The defects then act on the rational boundary states as
\begin{equation}
{\cal D}_{LMS=0} \braneket{[L',\CFTM'_{CFT}, S'=\pm 1]} = \sum N_{L,L'}^{L''} \braneket{[L'', M+M'_{CFT}, \pm 1} \; ,
\end{equation}
where $N$ are the $\mathfrak{su}(2)_k$ fusion rule coefficients. 

Likewise, the fusion of the defect lines follows the fusion rules for the supersymmetric minimal models. 


In particular, the defects labeled $(0,M,0)$, $M=0, 2, \dots, 2k+4$, realize the ${\mathbb Z}_{k+2}$ symmetry of the model in terms of defects. The action of the generator $(0,2,0)$ on the branes can be identified with a rotation by an angle of $2\pi/(k+2)$ on the branes in the Landau-Ginzburg picture. Higher $L$ defects act on single wedges in the LG picture by blowing them up to a union of $L+1$ wedges. 

Note that starting from the topological defects with $L=0$, we can build higher $L$ topological defects by forming "bound states", i.e. by  perturbing superpositions of lower $L$ defects by relevant defect fields, leading to a non-trivial flow on the defect. The pattern for the bound state formation is exactly the same as for the case of Cardy boundary states. This can again be checked on the level of the RR charges which add up when forming new bound states from superpositions, see \cite{Brunner:2005fv} for a discussion of this in the "folded" picture, where defects correspond to permutation branes.  Indeed, the charges of defects and boundary states only differ by a factor of $\sqrt{S_{0l}}$ which, for example, can be absorbed in a normalization factor of the vacuum so that on the level of representation labels the computations for the Cardy states and defects become the same.

This point of view will be useful for us in later sections, where we model the  topological defects on the mirror side, using matrix factorizations, as it allows a direct match starting from the symmetry defects.

\subsection{RG flows and CFT}


Let us comment on some special features. First of all, the perturbations we want to consider do not change any spin structures. Under the perturbation, NSNS sectors flow to NSNS sectors and RR sectors flow to RR sectors. Likewise, the boundary conditions (\ref{eq:Aglue})  on the supercurrents are left untouched by the perturbation. This means that the choice to pick $S$ odd for the spin structure of the branes is compatible with the perturbation and allows us to consistently match the Landau-Ginzburg picture of the branes with the concrete boundary states before and after the flows. Furthermore, the bulk perturbations we consider do not induce a flip of orientation on the branes. Note that the topological defect with representation labels $(0,0,2)$ acts on bulk states by assigning $1$ to states in the NS sector and $(-1)$ to states in the $R$ sector, and on branes by flipping their orientation. We can conclude  that the topological defect with labels $(0,0,2)$ is always preserved by the flows. 

We can now bring together the LG and CFT point of view. As described above, from the point of view of the LG model, RG flows can be described by specifying which subset of the basic branes decouples from the CFT part of the theory and becomes massive. From this, we can deduce that symmetry preserving flows require a consistent action of the preserved symmetries on the remaining branes. This means that the corresponding defects remain symmetry defects, and in particular topological, along the flow.

To perform a consistent and precise analysis of all topological defects, we will in the following invoke mirror symmetry. For minimal models, one can obtain the mirror  by an orbifold procedure. Modding out by ${\mathbb Z}_{k+2}$ brings us to the partition function
\begin{equation}
Z=\sum \chi_{l,m,s}(q) \bar\chi_{l,-m,s}(\bar{q}) \; .
\end{equation}
To study the A-branes of the original model, one has to consider the B-branes of the mirror theory. The set of topological defects that preserves the full supersymmetry remains invariant. 

For us, going to the mirror side has the advantage that the RG flows can be described very conveniently in terms of matrix factorizations. In this way, we control the part of the fusion that is protected by supersymmetry and infer how branes and defects get mapped under RG flows. We will do so in the following sections, recovering the surviving symmetries for the flows and extending the analysis to the preservation of higher $L$ topological defects.

\section{Review of matrix factorizations}\label{sec:MF}


\subsection{Properties of matrix factorizations}
B-type D-branes in Landau-Ginzburg models have a description in terms of matrix factorizations of the Landau-Ginzburg superpotential $W$ \cite{Kapustin:2002bi, Brunner:2003dc, Lazaroiu:2003zi}. The folding trick \cite{Bachas:2001vj} relates defects and boundary states so that B-type defects can be described by matrix factorizations of the difference $W = W_1 - W_2$ of the superpotentials of the theories on either side of the defect \cite{Brunner:2007qu}. In this sense, boundary states can be regarded as special cases of defects, \textit{i.e.} defects with a trivial theory on one side.
Usually, a matrix factorization $P$ of $W  \in  S$ is represented by
\begin{equation}\label{MF}
P  \; : \;\;\;  P_1 \; {{\underrightarrow{\;\;\;\;\;  p_1\;\;\;\;\;}} \atop {\overleftarrow{\;\;\;\;\;  p_0 \;\;\;\;\;}}} \; P_0  \; , \;\;\; p_1 p_0 = W \text{id}_{P_0} \; , \;\;\; p_0 p_1 = W \text{id}_{P_1} \; .
\end{equation}
$S$ denotes a polynomial ring over $\mathbb{C}$ in the chiral fields contained in $W$. The modules $P_i$ are $n$-dimensional free $S$-modules, and the maps $p_i$ are matrices of arbitrary but equal rank $n$ which factorize to $W$ times the identity map.
The data of a matrix factorization $P$ can be assembled in the odd matrix
\begin{equation}
d_P = \begin{pmatrix}
0 & p_1\\
p_0 & 0
\end{pmatrix} \; ,
\end{equation}
which specifies the boundary or defect contribution to the total BRST-charge of the topological model. Boundary or defect fields then correspond to the cohomology of the boundary or defect BRST-operator. Physically, the matrix $d_P$ can be thought of as containing the tachyon profile which forces a brane-anti-brane or defect-anti-defect pair in the underlying sigma model to form the brane or defect described by $P$.\\ 
Matrix factorizations can be regarded as two-periodic complexes which are twisted by $W$ and they form a triangulated additive category \cite{Orlov:2003yp}. The matrix factorizations corresponding to branes or defects constitute the objects of the category and topological string states constitute the morphisms between two matrix factorizations $P$ and $Q$. The space of topological string states is given by the cohomology of
\begin{equation}
\text{Hom} (P,Q) = \bigoplus_{i,j = 0,1} \text{Hom} (P_i, Q_j) \; ,
\end{equation}
\textit{i.e.} physical morphisms are required to be closed with respect to the differential $d$ and taken modulo those morphisms which are $d$-exact.
The differential acts on a morphism $\phi$ as
\begin{equation}
d \phi = d_Q \circ \phi - (-1)^{\text{deg}(\phi)} \phi \circ d_P \; ,
\end{equation}
where the $\mathbb{Z}_2$-grade of $\phi$ is given by $\text{deg}(\phi) = (i-j)\text{mod} \, 2$. Morphisms of degree zero are referred to as bosons. They consist of two components $b = (b_0, b_1)$ which map $b_i : P_i \rightarrow Q_i$ and are represented by an even matrix. Morphisms of degree one are referred to as fermions $f = (f_0, f_1)$. They map $f_i : P_i \rightarrow Q_{i+1}$ and are represented by an odd matrix. 
Turning on relevant perturbations between two initial objects, the combined system flows to a new object.
On the level of matrix factorizations, the perturbed object is described by the cone $\text{Cone} (f: P \rightarrow Q)$, which corresponds to the direct sum of the initial matrix factorizations, deformed by the tachyon:
\begin{equation}\label{DefCone}
P':= \text{Cone} (f: P \rightarrow Q) : \;\;\; P_1 \oplus Q_1 \; {{\underrightarrow{\;\;\;\;\; p'_1\;\;\;\;\;}} \atop {\overleftarrow{\;\;\;\;\; p'_0 \;\;\;\;\;}}} \; P_0 \oplus Q_0\; , 
\end{equation}
where
 \begin{equation}\label{MapsCone}
 p'_0 =
\begin{pmatrix}
  p_0 & 0 \\
  f_0 & q_0\\
 \end{pmatrix} \; ,\;\;\;\;\;
  p'_1 =
\begin{pmatrix}
  p_1 & 0 \\
  f_1 & q_1\\
 \end{pmatrix} \; .
 \end{equation}
A crucial ingredient for studying bound state formation is the concept of equivalences. Two matrix factorizations specified by $d_P$ and $d_Q$ are equivalent if block matrices $U$ and $V$ can be found which satisfy
\begin{equation}
U d_P V = d_Q \; , \;\;\; UV = \text{id}_Q + \{d_Q, O_Q\} \; , \;\;\;  VU = \text{id}_P + \{d_P, O_P\} \; ,
\end{equation}
for some $O_P$ and $O_Q$, \textit{i.e.} $U$ and $V$ are inverse up to BRST-trivial terms and the mapping preserves the structure of the corresponding spaces of morphisms.
Matrix factorizations for which the space of topological string states is empty are trivial.  Any trivial matrix factorization is equivalent to the rank one factorization specified by
\begin{equation}
d_{trivial} = \begin{pmatrix}
0 & 1 \\
W & 0
\end{pmatrix} \; . 
\end{equation}
Trivial matrix factorizations can be added to any other matrix factorization without changing the topological spectrum, \textit{i.e.} there is an equivalence
\begin{equation}
d_P \sim d_P \oplus d_{trivial} \; .
\end{equation}
A very important feature of the category of matrix factorizations is that it has adjoints. 
Bending an object of the topological model to the left or right results in a reversal of orientation. The oppositely oriented versions of an object described by a matrix factorization $P$ are given by its left and right adjoints, ${}^\dagger P$ and $P^\dagger$, respectively. Depending on the model, the two versions may differ. Adjunctions of defects have been studied in \cite{Carqueville:2012st, Carqueville:2013usa}.
For a B-type defect $P$ between Landau-Ginzburg models with superpotentials $W_1 \in \mathbb{C}[X_1,...,X_m]$ and $W_2 \in \mathbb{C}[Y_1,...,Y_n]$, they are given by
\begin{equation}
P^\dagger \simeq P^\vee [n] \; , \;\;\; {}^\dagger P \simeq P^\vee [m] \; ,
\end{equation}
where the dual $P^\vee$ of a matrix factorization $P$ is defined by 
\begin{equation}\label{DefDualMF}
P^\vee \; : \; P_1^\vee \; {{\underrightarrow{\;\;\; p_0^\vee \;\;\;}} \atop {\overleftarrow{\;\; - p_1^\vee \;\;}}} \; P_0^\vee \; ,
\end{equation}
and $[m]$ denotes a shift functor which shifts the $\mathbb{Z}_2$-grade by $m$. The matrix factorization $P[1]$ is defined by 
\begin{equation}\label{DefShiftFunctor}
P[1]  \; : \;\;\;  P_0 \; {{\underrightarrow{\;\;\;\;\; - p_0\;\;\;\;\;}} \atop {\overleftarrow{\;\;\;\;\; - p_1 \;\;\;\;\;}}} \; P_1  \; ,
\end{equation}
and physically corresponds to the anti-brane or anti-defect of the brane or defect described by $P$.\\
On the level of the topological model, supersymmetry preserving defects can be composed with other defects and with boundary states which preserve the same supersymmetry without having to deal with regularization. In terms of matrix factorizations, fusion of two B-type objects $P$ and $Q$ amounts to taking the graded tensor product $P' := P \otimes Q$ \cite{Brunner:2007qu, Ashok:2004zb} which is given by
\begin{equation}
P'_1 = (P_1 \otimes Q_0) \oplus (P_0 \otimes Q_1) \; {{\underrightarrow{\;\; p'_1\;\;}} \atop {\overleftarrow{\;\; p'_0 \;\;}}} \; (P_0 \otimes Q_0) \oplus (P_1 \otimes Q_1) = P'_0 \; ,
\end{equation}
where
\begin{equation}
p'_1 = \begin{pmatrix}
p_1 \otimes \text{id}_{Q_0} & - \text{id}_{P_0} \otimes q_1\\
\text{id}_{P_1} \otimes q_0 & p_0 \otimes \text{id}_{Q_1}
\end{pmatrix} \; , \;\;\;\;\; p'_0 = \begin{pmatrix}
p_0 \otimes \text{id}_{Q_0} &  \text{id}_{P_1} \otimes q_1\\
- \text{id}_{P_0} \otimes q_0 & p_1 \otimes \text{id}_{Q_1}
\end{pmatrix} \; .
\end{equation}
Let $P$ be a matrix factorization over $\mathbb{C}[X_i, Y_i]$, describing a defect between Landau-Ginzburg models with superpotentials $W_1 (X_i)$ and $W_2 (Y_i)$, and let $Q$ be a matrix factorization over $\mathbb{C}[Y_i, Z_i]$, representing a defect between the models with superpotentials $W_2 (Y_i)$ and $W_3 (Z_i)$. Then, $P'$ is actually a matrix factorization over $\mathbb{C}[X_i, Y_i, Z_i]$. However, bringing the two defects on top of each other results in a new defect between the models with superpotentials $W_1 (X_i)$ and $W_3 (Z_i)$. Consequently, $P'$ has to be regarded as a matrix factorization over $\mathbb{C}[X_i, Z_i]$. This can be achieved by expanding
\begin{equation}
\mathbb{C}[X_i, Y_i, Z_i] = \bigoplus\limits_{(\alpha_1, ..., \alpha_n) \in \mathbb{N}_0^n} Y_1^{\alpha_1} \cdot ... \cdot Y_n^{\alpha_n} \mathbb{C}[X_i, Z_i] \; ,
\end{equation}
which, however, gives infinite rank to $P'$. It has been shown \cite{Brunner:2007qu} that, provided $P$ and $Q$ are of finite rank, $P'$ can be reduced to finite rank by splitting off infinitely many trivial matrix factorizations. The reduced matrix factorization will be denoted by $P \ast Q$. On the level of modules, it can be obtained as the matrix factorization associated to the module
\begin{equation}\label{moduleM}
M = \text{coker}(p_1 \otimes \text{id}_{Q_0}, \text{id}_{P_0} \otimes q_1) \; ,
\end{equation}
regarded as a $\mathbb{C}[X_i, Z_i]$-module \cite{Brunner:2007qu}. Likewise, in case $Q$ describes a brane in the model with $W_2 (Y_i)$, the matrix factorization corresponding to the new boundary is associated to \eqref{moduleM} regarded as a $\mathbb{C}[X_i]$-module.\\

\noindent
The matrix factorization formalism carries over to orbifolds of Landau-Ginzburg models. In case there is a finite group $\Gamma$ which acts on the polynomial ring $S$ in such a way that the superpotential $W \in S$ of the Landau-Ginzburg model is invariant, it is possible to consider the corresponding $\Gamma$-orbifold. In the orbifold model, B-type branes can then be represented by $\Gamma$-equivariant matrix factorizations \cite{Ashok:2004zb}, \textit{i.e.} \eqref{MF} together with representations $\rho_i$ of $\Gamma$ on the modules $P_i$ which are compatible with the $S$-module structure and commute with the maps $p_i$:
\begin{equation}\label{compatibility}
\rho_i (\gamma) (s \cdot p) = \rho(\gamma)(s) \cdot \rho_i (\gamma) (p) \; , \;\;\; \rho_{i+1} (\gamma) p_i = p_i \rho_i (\gamma) \; ,
\end{equation}
where  $\gamma \in \Gamma$, $s \in S$, $p \in P_i$ and the action of $\Gamma$ on the polynomial ring $S$ is denoted by $\rho$. 
Defects between orbifolds of Landau-Ginzburg models with orbifold groups $\Gamma_1$ and $\Gamma_2$ are described by $\Gamma := \Gamma_1 \times \Gamma_2$-equivariant matrix factorizations \cite{Brunner:2007ur}. Often, $\Gamma$-equivariant matrix factorizations for defects can be found from matrix factorizations of the unorbifolded model using the orbifold construction procedure.
The first step is to find the subgroup $\Gamma' \subset \Gamma$ which stabilizes the map $p_1$ and fix a representation of $\Gamma'$ on the matrix factorization of the unorbifolded model. A $\Gamma$-equivariant matrix factorization can then be obtained by taking the sum of the $\Gamma / \Gamma'$-orbit of this $\Gamma'$-equivariant matrix factorization. 
The orbifold category is obtained by additionally imposing equivariance conditions on \eqref{DefCone}, \eqref{DefShiftFunctor}, as well as on the space of physical morphisms of the unorbifolded model \cite{Ashok:2004zb}.  
Likewise, the fusion product of two objects $P$ and $Q$ of the orbifold model can be found as the matrix factorization associated to the part $M^{\Gamma_{sqeezed}}$ of the module \eqref{moduleM} which is invariant with respect to the orbifold group $\Gamma_{sqeezed}$ of the squeezed-in model \cite{Brunner:2007ur}. The fusion product will be denoted by $P \ast_{orb} Q$. We refer to \cite{Carqueville:2012dk} for a concise discussion of consistent orbifold theories obtained as generalized orbifolds, assuming in particular the agreement of left and right adjoints.  \\

\noindent
For the single-variable model with superpotential $W = X^d$, multiplication of the field with $d$-th roots of unity leaves the superpotential invariant and is a symmetry of the theory. In particular, the symmetry action agrees with  the $\mathbb{Z}_d$-action identified in section \ref{sec:A-picture} and hence the following discussion allows to describe the topological sector of the  orbifold theory. We will denote the orbifold model by $X^d/\mathbb{Z}_d$.
An elementary matrix factorization for a boundary condition $\mathcal{B}^{(N,M)}$ in $X^d/\mathbb{Z}_d$ is  specified by an integer $N \in \{1,...,d-1\}$ and a $\mathbb{Z}_d$-representation label $M \in \mathbb{Z}_d$:
\begin{equation}\label{MFEquivBC}
\mathcal{B}^{(N,M)} (X) \; : \; \mathbb{C}[X][M+N] \; {{\underrightarrow{\;\;\;\; p_1= X^N \;\;\;\;}} \atop {\overleftarrow{\;\; p_0 = X^{d-N}\;\;}}} \; \mathbb{C}[X][M] \; ,
\end{equation}
where $[.]$ denotes the $\mathbb{Z}_d$-charge of $\mathbb{C} \subset \mathbb{C}[X]$. The matrix factorizations $\mathcal{B}^{(N,M)}$ have been identified \cite{Ashok:2004zb} with rational boundary states $\braneket{L, M_{CFT}, 1}$ in the corresponding minimal model orbifold $\mathcal{M}_{d-2}/\mathbb{Z}_d$ as
\begin{equation}\label{identificationBCs}
\braneket{N-1, 2M + N, 1} \leftrightarrow \mathcal{B}^{(N,M)} \; .
\end{equation}
A $\Gamma = \mathbb{Z}_{d_1} \times \mathbb{Z}_{d_2}$-equivariant matrix factorization $P$ for a defect between Landau-Ginzburg orbifolds $X^{d_1}/\mathbb{Z}_{d_1}$ and $Y^{d_2}/\mathbb{Z}_{d_2}$ takes the following general form:
\begin{equation}\label{GeneralEquivMF}
\tikzcdset{row sep/normal=0.7cm}
\tikzcdset{column sep/normal=0.7cm}
\begin{tikzcd}
P : S^{M} \begin{pmatrix}
\lbrack l_M, r_M \rbrack \\ \lbrack l_{M+1}, r_{M+1} \rbrack \\  \vdots \\ \lbrack l_{2M - 1} , r_{2M - 1}\rbrack \\
\end{pmatrix} \arrow[r, shift left, "p_{1}"]  & \arrow[l, shift left, "p_0"]  S^{M} \begin{pmatrix}
\lbrack l_0, r_0 \rbrack \\ \lbrack l_1, r_1 \rbrack \\  \vdots \\ \lbrack l_{M-1}, r_{M-1} \rbrack \\
\end{pmatrix}
\end{tikzcd} ,
\end{equation}
where $S = \mathbb{C}[X,Y]$ and $[.,.]$ denotes the $\mathbb{Z}_{d_1} \times \mathbb{Z}_{d_2}$-charges on $\mathbb{C} \subset \mathbb{C}[X,Y]$.
Matrix factorizations for adjoints of a defect $P$ defined by \eqref{GeneralEquivMF} have been determined in \cite{Klos:2019axh}.
The right adjoint $P^\dagger$ is given by
\begin{equation}\label{DefGeneralRightAdjoint}
\tikzcdset{row sep/normal=0.7cm}
\tikzcdset{column sep/normal=0.7cm}
\begin{tikzcd}
P^\dagger : S^{M} \begin{pmatrix}
\lbrack -r_0 + 1, - l_0 \rbrack \\ \lbrack -r_1 + 1, - l_1 \rbrack \\  \vdots \\ \lbrack  -r_{M - 1} + 1 , - l_{M - 1}\rbrack \\
\end{pmatrix} \arrow[r, shift left, "p_{1}^T"]  & \arrow[l, shift left, " - p_0^T"]  S^{M} \begin{pmatrix}
\lbrack - r_M + 1, - l_M \rbrack \\ \lbrack - r_{M+1} + 1, - l_{M+1} \rbrack \\  \vdots \\ \lbrack - r_{2M - 1} + 1, - l_{2M - 1} \rbrack \\
\end{pmatrix}
\end{tikzcd} \; ,
\end{equation}
where $[.,.]$ denotes the $\mathbb{Z}_{d_2} \times \mathbb{Z}_{d_1}$-charges on $\mathbb{C}$. The left adjoint ${}^\dagger P$ differs from $P^\dagger$ by a shift by $[-1, 1]$ in the $\mathbb{Z}_{d_2} \times \mathbb{Z}_{d_1}$-charges:
\begin{equation}\label{DefGeneralLeftAdjoint}
\tikzcdset{row sep/normal=0.7cm}
\tikzcdset{column sep/normal=0.7cm}
\begin{tikzcd}
{}^\dagger P : S^{M} \begin{pmatrix}
\lbrack -r_0 , - l_0 + 1 \rbrack \\ \lbrack -r_1 , - l_1+ 1 \rbrack \\  \vdots \\ \lbrack  -r_{M - 1}  , - l_{M - 1} + 1 \rbrack \\
\end{pmatrix} \arrow[r, shift left, "p_{1}^T"]  & \arrow[l, shift left, " - p_0^T"]  S^{M} \begin{pmatrix}
\lbrack - r_M , - l_M + 1 \rbrack \\ \lbrack - r_{M+1} , - l_{M+1} + 1 \rbrack \\  \vdots \\ \lbrack - r_{2M - 1} , - l_{2M - 1} + 1 \rbrack \\
\end{pmatrix}
\end{tikzcd} \; .
\end{equation}

\subsection{Matrix factorizations for topological defects}\label{sec:MFTopDefs}
Defects between single-variable Landau-Ginzburg models with superpotentials $W_1 = X^d$ and $W_2 = Y^d$ and symmetry groups $\Gamma_1 = \Gamma_2 = \mathbb{Z}_d$ are represented by matrix factorizations of $W=X^d - Y^d$. 
A special subclass of rank one matrix factorizations of $W$ is specified by
\begin{equation}\label{TopMFunorbifolded}
P_{\{\alpha, \alpha + 1, ..., \alpha + \mathcal{N} - 1\}} \; :	\;\;\; \mathbb{C}[X,Y] \; {{\underrightarrow{\;\; p_1 = \prod\limits_{i=0}^{\mathcal{N}-1} (X - \eta^{\alpha + i} Y)\;\;}} \atop {\overleftarrow{\;\; p_0 = \prod\limits_{i' = \mathcal{N}}^{d-1} (X - \eta^{\alpha + i'} Y) \;\;}}} \;  \mathbb{C}[X,Y]\; ,
\end{equation}
where $\alpha \in \mathbb{Z}_d$, $\mathcal{N} \in \{1,...,d-1\}$ and $\eta$ denotes a $d$-th root of unity. These matrix factorizations have been identified \cite{Brunner:2007qu} with topological defects $\mathcal{D}_{LM}$ in the corresponding IR fixed point CFT $\mathcal{M}_{d-2}$ as
\begin{equation}
P_{\{\alpha, \alpha + 1, ..., \alpha + \mathcal{N} - 1\}} \leftrightarrow \mathcal{D}_{\mathcal{N}-1, \mathcal{N}-1 + 2 \alpha} \; .
\end{equation}
The subclass of matrix factorizations with linear $p_1$, \textit{i.e.} $\mathcal{N}=1$, corresponds to the symmetry defects associated to the $\mathbb{Z}_d$-symmetry of the model. The linear factor encodes the gluing condition along the defect line. The defects $P_{\{\alpha\}}$ and $P_{\{d-\alpha\}}$ are inverse of one another and compose to the identity defect $P_{\{0\}}$.\\ 

\noindent
In the corresponding $\mathbb{Z}_d$-orbifold, defects are represented by $\Gamma = \mathbb{Z}_d \times \mathbb{Z}_d$-equivariant matrix factorizations, where the first $\mathbb{Z}_d$ acts only on $X$ and the second $\mathbb{Z}_d$ acts only on $Y$. 
Equivariant matrix factorizations for topological defects can be obtained from the non-equivariant objects \eqref{TopMFunorbifolded} by means of the orbifold construction. The subgroup $\Gamma' \subset \Gamma$ which leaves \eqref{TopMFunorbifolded} invariant is given by the diagonal subgroup $\Gamma' \simeq \mathbb{Z}_d$ which multiplies both fields with the same root of unity. By the factorization condition and \eqref{compatibility}, fixing the charge of $\mathbb{C} \subset P_0$ under $\Gamma'$ to be $\mathcal{M} \in \mathbb{Z}_d$ determines a $\Gamma'$-representation on the whole module $P_0$ and extends to a representation on the module $P_1$. 
The $\Gamma/\Gamma' \simeq \{1\} \times \mathbb{Z}_d$-orbit of this $\Gamma'$-equivariant matrix factorization then yields a $\Gamma$-equivariant matrix factorization $\tilde{P}^{(\mathcal{N}, \mathcal{M})}$:
\begin{equation}\label{EquivTopMFstandardbasis}
\tilde{p}_1^{(\mathcal{N}, \mathcal{M})} = \bigoplus\limits_{j \in \mathbb{Z}_d}  \Bigg[   \prod\limits_{i=0}^{\mathcal{N}-1} (X - \eta^{\alpha + i + j} Y) \Bigg]  \; : \; \Big( \mathbb{C}[X,Y] \; [\mathcal{M} + \mathcal{N}] \Big)^{\oplus d} \rightarrow \Big( \mathbb{C}[X,Y] \; [\mathcal{M}] \Big)^{\oplus d} \; ,
\end{equation}
where [.] denotes the $\mathbb{Z}_d$-charge of $\mathbb{C} \subset \mathbb{C}[X,Y]$ and we only specify the map $\tilde{p}_1$ as $\tilde{p}_0$ is then determined by the factorization condition. In the orbifold model, matrix factorizations with different values of $\alpha$ are equivalent so that $\alpha$ can be set to zero without loss of generality. Often, it is convenient to perform a basis change from the standard basis $e_i$, in which \eqref{EquivTopMFstandardbasis} is diagonal, to a basis $\tilde{e}_i$ in which the action of $\Gamma$ on \eqref{EquivTopMFstandardbasis} becomes diagonal. This change of basis is implemented by \cite{Brunner:2009zt}
\begin{equation}\label{BasisChange}
\tilde{e}_a = \sum\limits_{i = 1}^{d} \eta^{ia} e_i \; , \;\;\; e_b = \frac{1}{d} \sum_{i = 1}^{d} \eta^{ib} \tilde{e}_i \; ,
\end{equation}
where $a,b \in \mathbb{Z}_d $. In this basis, the $\Gamma$-equivariant matrix factorization $\mathcal{T}^{(\mathcal{N}, \mathcal{M})}$ for a topological defect reads
\begin{equation}\label{TopMF1}
\mathcal{T}^{(\mathcal{N}, \mathcal{M})} (X,Y) \; : \; S^d \begin{pmatrix}
[\mathcal{M} + \mathcal{N}, 0]\\
[\mathcal{M} + \mathcal{N} + 1, -1]\\
\vdots\\
[\mathcal{M} + \mathcal{N} + d - 1, -d + 1]
\end{pmatrix}  {{\underrightarrow{\;\;\;p_1^{(\mathcal{N}, \mathcal{M})} \;\;\;}} \atop {\overleftarrow{\;\;\; p_0^{(\mathcal{N},  \mathcal{M})}  \;\;\;}}} S^d \begin{pmatrix}
[\mathcal{M}, 0]\\
[\mathcal{M} + 1, -1]\\
\vdots\\
[\mathcal{M} + d - 1, - d + 1]
\end{pmatrix} \; ,
\end{equation}
where $[.,.]$ denotes the $\mathbb{Z}_d \times \mathbb{Z}_d $-action on the subspace $\mathbb{C} \subset \mathbb{C}[X,Y] =: S$, the first entry corresponding to the $\mathbb{Z}_d $ acting on $X$, the second one corresponding to the $\mathbb{Z}_d $ acting on $Y$. The map $p_1^{(\mathcal{N}, \mathcal{M})} (X,Y)$ is given by 
		\begin{equation}\label{TopMF}		
p_1^{(\mathcal{N}, \mathcal{M})} (X,Y) = \begin{pmatrix}
		 X^{\mathcal{N}} &  & & & & - \sigma_1^{(\mathcal{N}, \mathcal{M})} X^{\mathcal{N}-1} Y\\
		- \sigma_1^{(\mathcal{N}, \mathcal{M})} X^{\mathcal{N}-1} Y & \ddots & & & & \vdots\\
		\vdots & \ddots & & & & (-1)^{\mathcal{N}} \sigma_{\mathcal{N}}^{(\mathcal{N}, \mathcal{M})} Y^{\mathcal{N}}\\
		(-1)^{\mathcal{N}} \sigma_{\mathcal{N}}^{(\mathcal{N}, \mathcal{M})} Y^{\mathcal{N}} & & & & & 0\\
		0 & \ddots & & & & \vdots \\
		\vdots & & & & &  0\\
		0 & & & & &  X^{\mathcal{N}}\\
		\end{pmatrix}_{d \times d} \; ,
		\end{equation}		
and the prefactors $\sigma_i^{(\mathcal{N}, \mathcal{M})}$ in \eqref{TopMF} take the following form:		
\begin{equation}\label{TopMFPrefactors}
\sigma_i^{(\mathcal{N}, \mathcal{M})} = \sum_{\alpha_1 = 0}^{\mathcal{N} - i} \sum_{\alpha_2 = \alpha_1 + 1}^{\mathcal{N} - (i - 1)} \cdots  \sum_{\alpha_i = \alpha_{i-1} + 1}^{\mathcal{N}  - 1} \eta^{\sum_j \alpha_j} \; .
\end{equation}	
The defects $\mathcal{T}^{(1, \mathcal{M})}$ have been shown \cite{Brunner:2007ur} to generate the quantum $\mathbb{Z}_d$-symmetry of the Landau-Ginzburg orbifold. They compose according to the symmetry group,
\begin{equation}
\mathcal{T}^{(1, \mathcal{M})} \ast_{orb} \mathcal{T}^{(1, \mathcal{M}')} = \mathcal{T}^{(1, \mathcal{M} + \mathcal{M}')} \; ,
\end{equation}
and their action on other objects of the model is to shift the $\mathbb{Z}_d$-representation label by $\mathcal{M}$. The defects $\mathcal{T}^{(1, \mathcal{M})}$ and $\mathcal{T}^{(1, d - \mathcal{M})}$ are inverse of one another and compose to the identity defect $\mathcal{T}^{(1, 0)} =: \mathcal{I}_d$. We note that, by \eqref{DefGeneralRightAdjoint} and \eqref{DefGeneralLeftAdjoint}, left and right adjoints of topological defects $\mathcal{T}^{(\mathcal{N}, \mathcal{M})}$ are equivalent:
\begin{equation}\label{adjointTops}
(\mathcal{T}^{(\mathcal{N}, \mathcal{M})})^\dagger \simeq \mathcal{T}^{(\mathcal{N}, - \mathcal{M} - \mathcal{N} + 1)}  \simeq {}^\dagger (\mathcal{T}^{(\mathcal{N}, \mathcal{M})}) \; .
\end{equation}\\

\noindent
\textbf{Bound state formation of topological defects}\\
Applying the cone construction and using the equivalences of matrix factorizations, it can be shown that topological defects can be perturbed in such a way that the resulting bound state is topological again. In particular, the following flows are possible:
\begin{equation}\label{boundstatestopmfs}
\text{Cone} \big( T: \mathcal{T}^{(1, \mathcal{M})} \rightarrow \mathcal{T}^{(\mathcal{N}', \mathcal{M}')} \big) \simeq \mathcal{T}^{(\mathcal{N}' + 1, \mathcal{M})} \; ,
\end{equation}
where 
\begin{equation}\label{Mlabelboundstates}
\mathcal{M}' \in \{\mathcal{M} + 1, \mathcal{M}, \mathcal{M}-1, ..., \mathcal{M} - \mathcal{N}' + 2\} \; ,
\end{equation}
and $T$ denotes the tachyon matrix. To see this, it is convenient to first consider the initial $\Gamma$-equivariant matrix factorizations in the standard basis, where they are defined by \eqref{EquivTopMFstandardbasis}, with $\alpha$ set to zero without loss of generality.
A topological defect specified by $\tilde{p}_1^{(\mathcal{N}' + 1, \mathcal{M}'')}$ can be obtained as a bound state of the initial defects by turning on tachyons between the component matrix factorizations $\big[ \tilde{p}_1^{(1, \mathcal{M})} \big]_{ii}$ and $\big[ \tilde{p}_1^{(\mathcal{N}', \mathcal{M}')} \big]_{(i+1)(i+1)}$. As one of them is linear and they do not share common factors, the topological spectrum between the two component matrix factorizations is purely fermionic with cohomology representatives
\begin{equation}\label{fermionspectrum}
f_1 = X^b \; , \;\;\; f_0 = - X^b \cdot \prod\limits_{\alpha \in D \setminus \{i,...,i+\mathcal{N}'\}} (X - \eta^\alpha Y) \; , \;\;\; 0 \le b \le \mathcal{N}' - 1 \; ,
\end{equation}
where $D = \{0,...,d-1 \}$. We refer to \cite{Brunner:2005fv} for details on topological string spectra between matrix factorizations of the form \eqref{TopMFunorbifolded}. To find the maps which survive the orbifold projection, it is necessary to impose equivariance conditions on the morphisms of the unorbifolded model. Requiring \eqref{fermionspectrum} to commute with the action of the orbifold group,
\begin{equation}
f_1 (X) \gamma^{\mathcal{M} + 1} = \gamma^{\mathcal{M}'} f_1 (\gamma X) \; ,
\end{equation}
yields that the tachyons with $b=0$ survive the orbifold projection iff \eqref{Mlabelboundstates} is satisfied.
A $d \times d$-dimensional tachyon matrix can then be constructed as
\begin{equation}
(\tilde{f}_1)_{i,j} := \delta^{(d)}_{i, j-1} \; .
\end{equation}
Performing elementary matrix transformations and splitting off trivial summands, one finds that the map $\tilde{p}'_1$ associated to the bound state,
\begin{equation}
\tilde{p}'_1 = \begin{pmatrix}
\tilde{p}_1^{(1, \mathcal{M})} & 0 \\
\tilde{f}_1 & \tilde{p}_1^{(\mathcal{N}', \mathcal{M}')}
\end{pmatrix}_{2d \times 2d} \; ,
\end{equation}
is indeed equivalent to a matrix $\tilde{p}_1^{(\mathcal{N}' + 1, \mathcal{M}'')}$. In the basis $\tilde{e}_i$ given by \eqref{BasisChange}, the tachyon matrix becomes
\begin{equation}
(\hat{f}_1)_{i,j} = \eta^i \delta^{(d)}_{i,j} \; ,
\end{equation}
and the bound state is characterized by
\begin{equation}\label{boundstatenewbasis}
\hat{p}_1 = \begin{pmatrix}
p_1^{(1, \mathcal{M})} & 0 \\
\hat{f}_1 & p_1^{(\mathcal{N}', \mathcal{M}')}
\end{pmatrix}_{2d \times 2d} \; , \quad \hat{P}_0 = (\mathbb{C}[X,Y])^{2d} \begin{pmatrix}
[\mathcal{M}, 0]\\
\vdots\\
[\mathcal{M}+d-1, -d+1]\\
[\mathcal{M}', 0]\\
\vdots\\
[\mathcal{M}' + d - 1, -d+1]
\end{pmatrix} \; ,
\end{equation}
where the component matrices $p_1^{(1, \mathcal{M})}$ and $p_1^{(\mathcal{N}', \mathcal{M}')}$ are defined by \eqref{TopMF}. 
The tachyon matrix can be used to eliminate the upper left and lower right block to find
\begin{equation}
\hat{p}_1 \simeq \begin{pmatrix}
0 & p_1^{(\mathcal{N}' + 1, \mathcal{M})} \\
1 & 0
\end{pmatrix}_{2d \times 2d} \; .
\end{equation}
In particular, the necessary operations do not alter the grading so that the $\mathbb{Z}_d \times \mathbb{Z}_d$-degrees on the module $\hat{P}_0$ are still given as in \eqref{boundstatenewbasis}. Splitting off trivial summands together with the corresponding degrees then yields \eqref{boundstatestopmfs}. Note that the tachyon in \eqref{boundstatestopmfs} is actually given by $T = (\hat{f}_0, \hat{f}_1)$ and the bound state is defined by \eqref{DefCone} and \eqref{MapsCone}. However, by the factorization condition and \eqref{compatibility}, it suffices to consider the matrix $p_1$ and the orbifold action on the module $P_0$.\\

\noindent
\textbf{The match with conformal field theory}\\
\noindent
We propose the following identification of the matrix factorizations $\mathcal{T}^{(\mathcal{N}, \mathcal{M})}$ in $X^d/\mathbb{Z}_d$ with topological defects $\mathcal{D}_{LM}$ in the corresponding minimal model orbifold $\mathcal{M}_{d-2}/\mathbb{Z}_d$:
\begin{equation}\label{identificationTopsCFTMF}
\mathcal{D}_{\mathcal{N} - 1, 2\mathcal{M} + \mathcal{N} - 1} \leftrightarrow \mathcal{T}^{(\mathcal{N}, \mathcal{M})} \; .
\end{equation}
Assuming this identification to hold, the translation of \eqref{boundstatestopmfs} to conformal field theory reads
\begin{equation}
\mathcal{D}_{0, M_1} \oplus \mathcal{D}_{L_2, M_2} \rightarrow \mathcal{D}_{L_2 + 1, M_1 + L_2 + 1} \; ,
\end{equation}
where $M_2 = M_1 + L_2 - 2a$, and $a \in \{-1, 0, +1, ..., L_2 - 1\}$. For $a = -1$, \textit{i.e.} $\mathcal{M'} = \mathcal{M} + 1$, this is in agreement with the results of \cite{Brunner:2005fv} on bound state formation of permutation boundary states in the tensor product of minimal models.
Using \eqref{identificationBCs} and \eqref{identificationTopsCFTMF}, the translation of the CFT fusion rule for the composition of topological defects with boundary conditions and with other topological defects reads
\begin{equation}\label{fusionalgebraTopTop&TopBC}
\begin{split}
&\mathcal{T}^{(\mathcal{N}, \mathcal{M})} \ast_{orb} \mathcal{B}^{(N,M)} = \bigoplus_{N' = |N - \mathcal{N}| +1}^{\text{min} (N + \mathcal{N} - 1, 2d - N - \mathcal{N} - 1)} \mathcal{B}^{(N', M')} \; ,\\
&\mathcal{T}^{(\mathcal{N}, \mathcal{M})} \ast_{orb} \mathcal{T}^{(\tilde{\mathcal{N}},\tilde{\mathcal{M}})} = \bigoplus_{\mathcal{N}' = |\tilde{\mathcal{N}} - \mathcal{N}| +1}^{\text{min} (\tilde{\mathcal{N}} + \mathcal{N} - 1, 2d - \tilde{\mathcal{N}} - \mathcal{N} - 1)} \mathcal{T}^{(\mathcal{N}', \mathcal{M}')} \; ,
\end{split}
\end{equation}
where
\begin{equation}
M' = \mathcal{M} + M + \frac{\mathcal{N} + N - N' - 1}{2} \; , \;\;\; \mathcal{M}' = \mathcal{M} + \tilde{\mathcal{M}} + \frac{\mathcal{N} + \tilde{\mathcal{N}} - \mathcal{N}' - 1}{2} \; , 
\end{equation}
and the sum in \eqref{fusionalgebraTopTop&TopBC} is taken in steps of two. By \eqref{adjointTops}, the fusion product of equivariant matrix factorizations for topological defects with their oppositely oriented versions then indeed contains the identity defect.

\subsection{Matrix factorizations for RG defects} \label{sec:RGdefects}
$\mathbb{Z}_d$-orbifolds of superconformal minimal models $\mathcal{M}_{d-2}$ admit $d-1$ relevant twisted chiral perturbations which induce an RG flow to some infrared orbifold $\mathcal{M}_{d'-2}/\mathbb{Z}_{d'}$ with $d' < d$. The perturbations preserve B-type supersymmetry and can be described in terms of B-type defects on the level of the corresponding Landau-Ginzburg orbifold. $\mathbb{Z}_{d'}\times \mathbb{Z}_d$-equivariant matrix factorizations of $W = X^{d'} - Y^d$ for RG defects describing RG flows from a UV model $Y^d/\mathbb{Z}_d$ to an IR model $X^{d'}/\mathbb{Z}_{d'}$ have been constructed in \cite{Brunner:2007ur}. They are determined by irreducible representations $m \in \mathbb{Z}_{d}$ and a $d'$-tuple of integers $n = (n_0, ..., n_{d' - 1})$ with $n_i \in \mathbb{N}_0$ such that $\sum_{i \in \mathbb{Z}_{d'}} n_i = d$. The matrix factorizations read:
\begin{equation}\label{RGMF}
\tikzcdset{row sep/normal=0.7cm}
\tikzcdset{column sep/normal=0.7cm}
\begin{tikzcd}
\mathcal{R}^{(m,n)} (X,Y): S^{d'} \begin{pmatrix}
\lbrack 1, -m \rbrack \\ \lbrack 2, -m - n_1 \rbrack \\  \vdots \\ \lbrack d', - m - \sum\nolimits_{i=1}^{d'-1} n_i \rbrack \\
\end{pmatrix} \arrow[r, shift left, "p_{1}^{(m,n)}"]  & \arrow[l, shift left, "p_0^{(m,n)}"]  S^{d'} \begin{pmatrix}
\lbrack 0, -m \rbrack \\ \lbrack 1, -m - n_1 \rbrack \\  \vdots \\ \lbrack d' - 1, - m - \sum\nolimits_{i=1}^{d'-1} n_i \rbrack \\
\end{pmatrix}
\end{tikzcd} ,
\end{equation}
where $S=\mathbb{C}[X,Y]$, $[.,.]$ specifies the $\mathbb{Z}_{d'} \times \mathbb{Z}_{d}$-action on $\mathbb{C} \subset \mathbb{C} [X,Y]$, and the map $p_1$ is given by
\begin{equation}\label{RGMF2}
p_1^{(m,n)} (X,Y) = \begin{pmatrix}
X  &  & &   - Y^{n_0}\\
- Y^{n_1} & \ddots &  							\\
 & \ddots & \ddots &            \\
 &  &     - Y^{n_{d'-1}} & X
\end{pmatrix}_{d' \times d'} \; .
\end{equation}
Right and left adjoint of an RG defect $\mathcal{R}^{(m,n)}(X,Y)$ are then represented by matrix factorizations of $Y^d - X^{d'}$ and describe a flow in the inverse direction via the same RG trajectory. They are given by 
\begin{equation}\label{MFRightAdjointRG}
\mathcal{R}^{(m,n)\dagger} \; : \;  S^{d'} \begin{pmatrix}
[m + 1, 0]\\
[m + 1 + n_1, -1]\\
\vdots\\
[m + 1 + \sum_{i=1}^{d' - 1} n_i, - d' + 1]
\end{pmatrix}  {{\underrightarrow{\; (p_1^{(m, n)})^T \;}} \atop {\overleftarrow{- (p_0^{(m, n)})^T }}} S^{d'} \begin{pmatrix}
[m + 1, -1]\\
[m + 1 + n_1, -2]\\
\vdots\\
[m + 1 + \sum_{i=1}^{d' - 1} n_i, 0]
\end{pmatrix} \; ,
\end{equation}
and 
\begin{equation}\label{MFLeftAdjointRG}
{}^\dagger\mathcal{R}^{(m,n)}: S^{d'} \begin{pmatrix}
[m , 1]\\
[m  + n_1, 0]\\
\vdots\\
[m + \sum_{i=1}^{d' - 1} n_i, - d' + 2]
\end{pmatrix}  {{\underrightarrow{\; (p_1^{(m, n)})^T \;}} \atop {\overleftarrow{ -(p_0^{(m, n)})^T  }}} S^{d'} \begin{pmatrix}
[m , 0]\\
[m + n_1, -1]\\
\vdots\\
[m  + \sum_{i=1}^{d' - 1} n_i, - d' +1]
\end{pmatrix} \; ,
\end{equation}
where $[.,.]$ denotes the $\mathbb{Z}_{d} \times \mathbb{Z}_{d'}$-action on $\mathbb{C} \subset \mathbb{C}[X,Y] = S$ and 
\begin{equation}\label{MapAdjointsRG}
(p_1^{(m,n)}(X,Y))^T = \begin{pmatrix}
X & -Y^{n_1} & & \\
& X & \ddots & \\
& & \ddots & \\
& & & \; \; \; -Y^{n_{d'-1} } \\
-Y^{n_0} & & & X\\
\end{pmatrix}_{d' \times d'} \; .
\end{equation}
For the special case that $m= 0$ and $n_i = 1 \; \forall i$, \textit{i.e.} $d = d'$, the RG defect reduces to the identity defect $\mathcal{T}^{(1,0)} =: \mathcal{I}_d$ in $X^d/\mathbb{Z}_d$, which is self-adjoint. Recall the fusion properties of \cite{Klos:2019axh} discussed in section \ref{sec:DefRG}. Composing RG defects with their adjoints yields the identity defect in the IR model:
\begin{equation}\label{RRdagger=id}
\mathcal{R}^{(m,n)} \ast_{orb} (\mathcal{R}^{(m,n)})^\dagger = \mathcal{I}_{d'} \; , \;\;\; \mathcal{R}^{(m,n)} \ast_{orb} {}^\dagger(\mathcal{R}^{(m,n)}) = \mathcal{I}_{d'} \; .
\end{equation}
Fusion in opposite order then yields a projection defect $\mathcal{P}^{(m,n)}$ in the UV model which projects onto the degrees of freedom of the IR model:
\begin{equation}\label{RdaggerR=P}
(\mathcal{R}^{(m,n)})^\dagger \ast_{orb} \mathcal{R}^{(m,n)} = \mathcal{P}^{(m,n)} \; , \;\;\;
{}^\dagger (\mathcal{R}^{(m,n)}) \ast_{orb} \mathcal{R}^{(m,n)} = {}^\dagger (\mathcal{P}^{(m,n)}) \; .
\end{equation}

\section{Results on topological defects in minimal models}\label{sec:Results}



In this section, we want to use the matrix factorization formalism for B-type defects and boundary conditions in $\mathbb{Z}_d$-orbifolds of single-variable Landau-Ginzburg models with superpotential $W = X^d$ to study the effect of perturbations of the bulk theory on topological defects, making the program outlined in section \ref{sec:DefRG} explicit in an example. Here, we use that the topological sector preserving B-type SUSY is under complete control. We collect the explicit description of the main ingredients from section \ref{sec:MF}, in particular, the  explicit description of the RG defects and their adjoints from section \ref{sec:RGdefects} as well as the description of topological defects from \ref{sec:MFTopDefs}.  Furthermore, all boundary and defect operations, in particular the fusion product, are under  control in this setting.

 
\subsection{Setup and general structures}

\begin{figure}[h]
\centering
\begin{tikzpicture}[thick]
\fill[fill=red!20!white] (3,0) -- (6,0) -- (6,3) -- (3,3) -- cycle;
\fill[fill=blue!20!white] (6,0) -- (10,0) -- (10,3) -- (6,3) -- cycle;
\fill[fill=red!20!white] (10,0) -- (13,0) -- (13,3) -- (10,3) -- cycle;
\draw[->, dashed] (6,1.5)--(7.9,1.5);
\draw[->, dashed] (10,1.5)--(8.1,1.5);
  \node[text width=1cm, anchor=west, right] at (3.5,0.5)
    {$X^{d'}/\mathbb{Z}_{d'}$};
    \node[text width=1cm, anchor=west, right] at (4,2.5)
    {IR};
    \draw[->, thick] (6,0) -- (6,1);
    \draw[->, thick] (6,1) -- (6,2);
    \draw[thick] (6,2) -- (6,3);
    \node[text width=1cm, anchor=west, right] at (6.3,0.5)
    {$Y^{d}/\mathbb{Z}_{d}$};
     \node[text width=1cm, anchor=west, right] at (6.5,2.5)
    {UV};
     \draw[->, thin] (8,0) -- (8,1);
     \draw[->, thin] (8,1) -- (8,2);
    \draw[thin] (8,2) -- (8,3);
     \node[text width=1cm, anchor=west, right] at (8.3,0.5)
    {$U^{d}/\mathbb{Z}_{d}$};
     \node[text width=1cm, anchor=west, right] at (8.5,2.5)
    {UV};
      \draw[->, thick] (10,3) -- (10,2);
       \draw[->, thick] (10,2) -- (10,1);
    \draw[thick] (10,1) -- (10,0);
     \node[text width=1cm, anchor=west, right] at (10.5,0.5)
    {$Z^{d'}/\mathbb{Z}_{d'}$};
     \node[text width=1cm, anchor=west, right] at (11,2.5)
    {IR};
   \node[text width=4cm, anchor=west, right] at (4.8,-0.4)
    {$\mathcal{R} (X,Y)$};
    \node[text width=4cm, anchor=west, right] at (7,-0.4)
    {$\mathcal{T}_{UV} (Y,U)$};
    \node[text width=4cm, anchor=west, right] at (9.1,-0.4)
    {$(\mathcal{R} (Z,U))^\dagger$};
\draw[->] (13.5,1.5) -- (14.5,1.5);
\fill[fill=red!20!white] (15,0) -- (18,0) -- (18,3) -- (15,3) -- cycle;
\draw[->, thin] (16.5, 0) -- (16.5,1);
\draw[->, thin] (16.5,1) -- (16.5,2);
\draw[thin] (16.5, 2) -- (16.5,3);
 \node[text width=1cm, anchor=west, right] at (15,0.5)
    {$X^{d'}/\mathbb{Z}_{d'}$};
    \node[text width=1cm, anchor=west, right] at (15.5,2.5)
    {IR};
\node[text width=1cm, anchor=west, right] at (16.5,0.5)
    {$Z^{d'}/\mathbb{Z}_{d'}$};
     \node[text width=1cm, anchor=west, right] at (17,2.5)
    {IR};
 \node[text width=4cm, anchor=west, right] at (15.5,-0.4)
    {$D_{IR} (X,Z)$};
\end{tikzpicture}
\caption{Setup for the RG flow of topological defects.}\label{FigureRGflowTopdefs}
\end{figure}
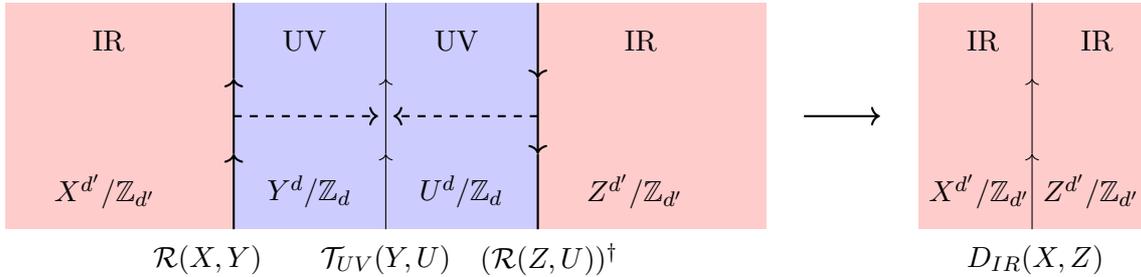

To study the fate of a topological defect $\mathcal{T}_{UV}$ in the initial UV Landau-Ginzburg orbifold $X^d/\mathbb{Z}_d$ under bulk perturbations, we perturb the UV model on both sides of the defect by the same relevant local operator, but keep the perturbation away from the defect. The RG flow then drives the perturbed regions to some IR model $X^{d'}/\mathbb{Z}_{d'}$ with $d'<d$, creating an RG interface $\mathcal{R}$ to the left of the topological defect and its left or right adjoint, ${}^\dagger \mathcal{R}$ or $\mathcal{R}^\dagger$, to the right. The resulting setup is depicted in figure \ref{FigureRGflowTopdefs} and has been discussed for general theories in section \ref{sec:DefRG}.
Taking the fusion product $\mathcal{R} \ast_{orb} \mathcal{T}_{UV} \ast_{orb} \mathcal{R}^\dagger$ then yields the defect $D_{IR}$ in the infrared model to which the topological defect $\mathcal{T}_{UV}$ flows. We are interested in pairs satisfying (\ref{eq:flowtop}) in the specific example:
\begin{equation}\label{TopFlowtoIR}
\mathcal{R} \ast_{orb} \mathcal{T}_{UV} \ast_{orb} \mathcal{R}^\dagger  = \mathcal{T}_{IR}\; ,
\end{equation}
as well as those that satisfy the  stronger intertwining property 
\begin{equation}\label{intertwining}
\mathcal{T}_{IR} \ast_{orb} \mathcal{R} = \mathcal{R} \ast_{orb} \mathcal{T}_{UV} \ast_{orb} \mathcal{P} = \mathcal{R} \ast_{orb} \mathcal{T}_{UV}\; .
\end{equation}

\subsection{Flows and lifts of branes}

To prepare our arguments, we work out the action of the RG defects and their adjoints on the branes. We give a pictorial description for both of them.\\

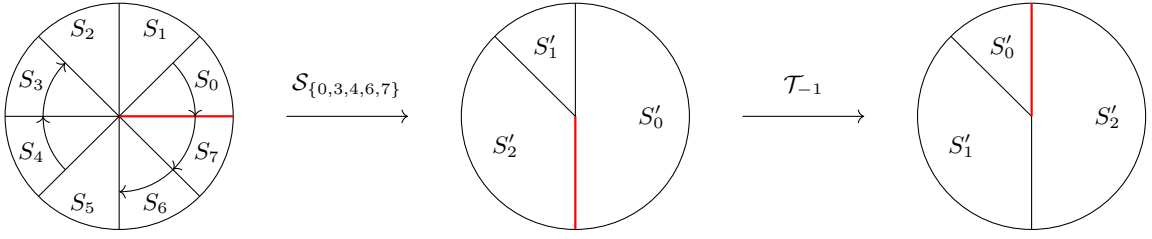
\begin{figure}[h]
\centering
 \begin{tikzpicture}
\def\R{1.5cm}
\draw (0,0) circle[radius=\R];
\draw (0,0) -- (0:\R)    (0,0) -- (45:\R)     (0,0) -- (90:\R)   (0,0) -- (135:\R)    (0,0) -- (180:\R)     (0,0) -- (225:\R)   (0,0) -- (270:\R) (0,0) -- (315:\R);
\draw[red, thick] (0,0) -- (0:\R);
\node[font=\footnotesize] at ( 22.5:\R/1.2) {$S_0$};
\node[font=\footnotesize] at (67.5:\R/1.2) {$S_1$};
\node[font=\footnotesize] at (112.5:\R/1.2) {$S_2$};
\node[font=\footnotesize] at (157.5:\R/1.2) {$S_3$};
\node[font=\footnotesize] at (202.5:\R/1.2) {$S_4$};
\node[font=\footnotesize] at (247.5:\R/1.2) {$S_5$};
\node[font=\footnotesize] at (292.5:\R/1.2) {$S_6$};
\node[font=\footnotesize] at (337.5:\R/1.2) {$S_7$};
\draw [->] ([shift=(180:1cm)]0,0) arc (180:135:1cm);
\draw [->] ([shift=(225:1cm)]0,0) arc (225:180:1cm);
\draw [->] ([shift=(315:1cm)]0,0) arc (315:270:1cm);
\draw [->] ([shift=(360:1cm)]0,0) arc (360:315:1cm);
\draw [->] ([shift=(45:1cm)]0,0) arc (45:0:1cm);
\draw[->] (2.2,0) -- (3.8,0);
\node[font=\footnotesize] at (3,0.4) {$\mathcal{S}_{\{0,3,4,6,7\}}$};
\draw (6,0) circle[radius=\R];
\draw  (6,0) --++ (90:\R)   (6,0) --++ (135:\R)      (6,0) --++ (270:\R) ;
\draw[red, thick] (6,0) --++ (270:\R);
\path ([shift=(0:1cm)]6,0)node[font=\footnotesize]{$S'_0$};
\path ([shift=(112.5:1cm)]6,0)node[font=\footnotesize]{$S'_1$};
\path ([shift=(202.5:1cm)]6,0)node[font=\footnotesize]{$S'_2$};
\draw[->] (8.2,0) -- (9.8,0);
\node[font=\footnotesize] at (9,0.4) {$\mathcal{T}_{-1}$};
\draw (12,0) circle[radius=\R];
\draw  (12,0) --++ (90:\R)   (12,0) --++ (135:\R)      (12,0) --++ (270:\R) ;
\draw[red, thick] (12,0) --++ (90:\R);
\path ([shift=(0:1cm)]12,0)node[font=\footnotesize]{$S'_2$};
\path ([shift=(112.5:1cm)]12,0)node[font=\footnotesize]{$S'_0$};
\path ([shift=(202.5:1cm)]12,0)node[font=\footnotesize]{$S'_1$};
\end{tikzpicture}
\caption{Pictorial representation of the action of $\mathcal{R}^{(2, (1,3,4))}$ on boundary conditions. The flow maps $\mathcal{B}_{UV}^{(1,2)} \mapsto \mathcal{B}_{IR}^{(1,0)} ,  \mathcal{B}_{UV}^{(1,5)} \mapsto \mathcal{B}_{IR}^{(1,1)} $, and  $\mathcal{B}_{UV}^{(1,1)} \mapsto \mathcal{B}_{IR}^{(1,2)}$. All other boundary conditions are annihilated.}\label{ExamplePictorialRepres}
\end{figure}

\noindent
\textbf{Bulk induced boundary flows of UV branes}\\
\noindent
For the model at hand, the branes and flows were addressed from the perspective of the A-model in section \ref{sec:A-picture}. Here, we address the question from the mirror B-type perspective following \cite{Brunner:2007ur}, to which we refer for further details.
In a first step, we briefly review a pictorial representation for the effect of bulk perturbations on boundary conditions of the UV theory. \\
\noindent 
Equivariant matrix factorizations for B-type boundary conditions in the Landau-Ginzburg orbifold $X^d/\mathbb{Z}_d$ are defined by \eqref{MFEquivBC}. For RG defects $\mathcal{R}^{(m,n)}$ defined by \eqref{RGMF} with $n_i \ge 1 \; \forall i \in \mathbb{Z}_{d'}$, the general fusion formula of \cite{Brunner:2007ur} implies that a boundary condition $\mathcal{B}^{(1, M_{UV})}$ is either annihilated or mapped to a boundary condition $\mathcal{B}^{(1,M_{IR})}$ with $M_{IR} \in \mathbb{Z}_{d'}$ in the IR model $X^{d'} / \mathbb{Z}_{d'}$. The boundary conditions surviving the flow are those with
\begin{equation}
M_{UV} \in m + \big \{0, n_1, n_1 + n_2, ..., \sum\nolimits_{i=1}^{d'-1} n_i  \big \} =: \mathcal{L}_{(m,n)} \; ,
\end{equation}
and the brane with $M_{UV} = m + \sum_{i=1}^{a} n_i$, where $a \in \mathbb{Z}_{d'}$, is mapped to the IR-brane with $M_{IR} = a$. The boundary conditions with $M_{UV} \notin \mathcal{L}_{(m,n)}$ are annihilated under the flow:
\begin{equation}\label{R*BC}
\mathcal{R}^{(m,n)} \ast_{orb} \mathcal{B}^{(1, M_{UV})} = \begin{cases} 0 \quad \;\;\;\;\;\;\;\;\;\;\;\;\;\; \text{for} \quad M_{UV} \in \mathcal{L}^c_{(m,n)}\\ \mathcal{B}^{(1, a)} \quad \;\;\;\;\;\;\;  \text{for} \quad M_{UV} = m + \sum_{i=1}^{a} n_i \; , \end{cases} 
\end{equation}
where $\mathcal{L}^c_{(m,n)}$ denotes the complement of the set $\mathcal{L}_{(m,n)}$. The formula \eqref{R*BC} can be depicted by considering a disk subdivided into $d$ wedges, each wedge representing a boundary condition with $N_{UV}=1$ in the model $X^d/\mathbb{Z}_d$. We mark one of the lines bounding the wedges and denote the sectors by $S_0,...,S_{d-1}$, starting from the marked line and going in counterclockwise direction. The boundary condition $\mathcal{B}^{(1, M_{UV})}$ is represented by the wedge $S_{M_{UV}}$. 
We define two operations on the disk. The operation $\mathcal{T}_{- \alpha}$ shifts the marked line by $\alpha$ steps in counterclockwise direction. The operation $\mathcal{S}_{\{\alpha_1, ..., \alpha_{d-d'}\}}$ annihilates the wedges $S_{\alpha_i}$ by merging the lines bounding them.
The action of an RG defect $\mathcal{R}^{(m,n)}$ on boundary conditions of the UV model can then be described by the following operation:
\begin{equation}
\mathcal{O}^{(m,n)} = \mathcal{S}_{\mathcal{L}^c_{(m,n)}-m} \mathcal{T}_{-m} = \mathcal{T}_{-a_{(m,n)}} \mathcal{S}_{\mathcal{L}^c_{(m,n)}}\; ,
\end{equation}
where $a_{(m,n)}$ denotes the number of wedges before the $m$-th wedge which are not annihilated, $a_{(m,n)}:= |\{0,...,m\} \cap \mathcal{L}_{(m,n)}\setminus \{m \}|$. The operation $\mathcal{O}^{(m,n)}$ then yields a disk subdivided into $d'$ wedges $S'_{M_{IR}}$, each wedge representing a boundary condition with $N_{IR}=1$ in the model $X^{d'}/\mathbb{Z}_{d'}$, see figure \ref{ExamplePictorialRepres} for an example. The pictorial representation for the action of RG defects can then be generalized to boundary conditions with $N_{UV}>1$ by representing $\mathcal{B}^{(N_{UV}, M_{UV})}$ by the union $S_{M_{UV}} \cup S_{M_{UV}+1} \cup ... \cup S_{M_{UV} + N_{UV} - 1}$ of wedges. Besides, the pictorial representation applies to RG defects with some $n_i = 0$ as well. In that case, the operation $\mathcal{S}$ is replaced by an operation which deletes $n_i - 1$ wedges for each $i$ with $n_i > 1$ and creates a new wedge for each $i$ with $n_i = 0$ by splitting wedges.
\\

\noindent
\textbf{Lifting branes from IR to UV}

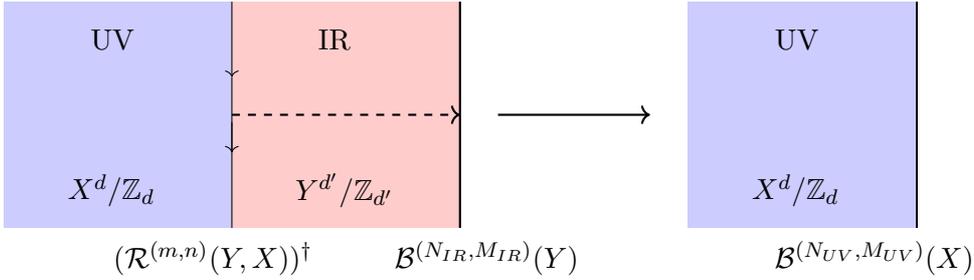
\begin{figure}[h]
\centering
\begin{tikzpicture}[thick]
\fill[fill=blue!20!white] (3,0) -- (6,0) -- (6,3) -- (3,3) -- cycle;
\fill[fill=red!20!white] (6,0) -- (9,0) -- (9,3) -- (6,3) -- cycle;
\draw[->, dashed] (6,1.5) -- (9,1.5);
  \node[text width=1cm, anchor=west, right] at (3.7,0.5)
    {$X^{d}/\mathbb{Z}_{d}$};
    \node[text width=1cm, anchor=west, right] at (4,2.5)
    {UV};
  \begin{scope}[thin]
    \draw[->] (6,3) -- (6,2);
    \draw[->] (6,2) -- (6,1);
    \draw (6,1.4) -- (6,0);
    \node[text width=1cm, anchor=west, right] at (6.7,0.5)
    {$Y^{d'}/\mathbb{Z}_{d'}$};
    \node[text width=1cm, anchor=west, right] at (7,2.5)
    {IR};
  \end{scope}
   \draw (9,0) -- (9,3);
   \node[text width=3.5cm, anchor=west, right] at (4.3,-0.4)
    {$(\mathcal{R}^{(m,n)}(Y,X))^\dagger$};
    \node[text width=3cm, anchor=west, right] at (8.0,-0.4)
    {$\mathcal{B}^{(N_{IR}, M_{IR})} (Y)$};
  \draw[->] (9.5,1.5) -- (11.5,1.5);  
  \draw[very thick] (15,0) -- (15,3);
 \fill[fill=blue!20!white] (12,0) -- (15,0) -- (15,3) -- (12,3) -- cycle;
  \node[text width=1cm, anchor=west, right] at (12.7,0.5)
    {$X^{d}/\mathbb{Z}_{d}$};
    \node[text width=1cm, anchor=west, right] at (13,2.5)
    {UV};   
\node[text width=3cm, anchor=west, right] at (13.0,-0.4)
    {$\mathcal{B}^{(N_{UV}, M_{UV})} (X)$};  
\end{tikzpicture}
\caption{Setup for the composition $\mathcal{R}^\dagger \ast_{orb} \mathcal{B}_{IR} = \mathcal{B}_{UV}$.}\label{FigureActionRdaggeronbc}
\end{figure}

\noindent
Likewise, we can describe the lift of the IR branes to the UV theory by merging the left or right adjoint of the RG interface with a boundary condition of the IR model. To do so, we consider the setup depicted in figure \ref{FigureActionRdaggeronbc}.
First, we consider the action of the right adjoint of the RG defect $\mathcal{R}^{(m,n)} (Y,X)$. We define $P:=(\mathcal{R}^{(m,n)}(Y,X))^\dagger$ and $Q:= \mathcal{B}^{(N_{IR}, M_{IR})} (Y)$. The matrix factorization for the right adjoint is defined by \eqref{MFRightAdjointRG} and \eqref{MapAdjointsRG}. The matrix factorization for the boundary condition in the IR model is defined by \eqref{MFEquivBC} with $M_{IR} \in \mathbb{Z}_{d'}$ and $N_{IR} \in \{1,..., d'-1\}$.
The generators $(e_a^P)_{a \in \mathbb{Z}_{d'}}$ of the module $P_0$ are of $\mathbb{Z}_d \times \mathbb{Z}_{d'}$-degree $[e_a^P] = [m+1+\sum_{i=1}^{a} n_i, -1-a]$ and the generator $e^Q$ of $Q_0$ carries $\mathbb{Z}_{d'}$-charge $[e^Q] = [M_{IR}]$. We denote the generators of $P_0 \otimes Q_0$ by $e_a := e_a^P \otimes e^Q$. Considered as a $\mathbb{C}[X]$-module, $P_0 \otimes Q_0$ is generated by $e_a^j := Y^j e_a$ with $j \in \mathbb{N}_0$. The basis vectors $e^j_a$ are of $\mathbb{Z}_d \times \mathbb{Z}_{d'}$-degree
\begin{equation}\label{ZZdegreeRdaggeronbc}
[e_a^j] = \big[m+1+ \sum\nolimits_{i=1}^{a} n_i , - 1 - a + M_{IR} + j \big] \; .
\end{equation} 
In this basis, the relations obtained from the module \eqref{moduleM} read
\begin{equation}\label{Rdaggeronbcrel1}
e^{j+1}_a = X^{n_a} e^j_{a-1} \; , \;\;\; e^{j+N_{IR}}_a = 0 \; .
\end{equation}
The first relation implies
\begin{equation}\label{Rdaggeronbcrel3}
e^j_a = X^{\sum_{i=0}^{j-1} n_{a-i}} e^0_{a-j} \; ,
\end{equation}
and can be used to eliminate all the basis vectors with $j>0$ from the generating system of the module $M$. Combining \eqref{Rdaggeronbcrel3} and and the second relation in \eqref{Rdaggeronbcrel1} yields
\begin{equation}\label{Rdaggeronbcrel4}
X^{\sum_{i=0}^{j+N_{IR}-1} n_{a-i}} e^0_{a-j-N_{IR}} = 0 \; .
\end{equation}			
The relations for $j>0$ follow from those for $j=0$ and the relations on the basis vectors $e^0_{a}$ then read
\begin{equation}\label{Rdaggeronbcrel5}
X^{\sum_{i=0}^{N_{IR}-1} n_{a + N_{IR} -i}} e^0_{a} = 0 \; .
\end{equation}
$\mathbb{Z}_{d'}$-invariance singles out the basis vector with $a = M_{IR} - 1$, which we denote by $f := e^0_{M_{IR}-1}$. It is of $\mathbb{Z}_d$-degree
\begin{equation}\label{ZdegreeRdaggeronbc}
[f] = \big[m+1+\sum\nolimits_{i=1}^{M_{IR}-1}n_i \big] \; ,
\end{equation}
and subject to the relation
\begin{equation}\label{Rdaggeronbcrel6}
X^{\sum_{i=0}^{N_{IR}-1} n_{M_{IR} - 1 + N_{IR} -i}} f = 0 \; .
\end{equation}
Consequently, the $\mathbb{Z}_{d'}$-invariant part $M^{\mathbb{Z}_{d'}}$ of the module $M$ is isomorphic to the module $\text{coker} \big(p_1^{(N_{UV}, M_{UV})}\big)$ associated to a boundary condition in the UV model and the fusion product reads
\begin{equation}\label{resultRdaggeronbc}
(\mathcal{R}^{(m,n)})^\dagger \ast_{orb} \mathcal{B}^{(N_{IR}, M_{IR})} = \mathcal{B}^{(N_{UV}, M_{UV})} \; ,
\end{equation}
with $M_{UV} \in \mathbb{Z}_d$ and $N_{UV} \in \{1,...,d-1\}$ given by 
\begin{equation}
M_{UV} = m + 1 + \sum\nolimits_{i=1}^{M_{IR}-1} n_i \quad , \quad N_{UV} = \sum\nolimits_{i=0}^{N_{IR}-1} n_{M_{IR} + N_{IR} -1-i}  \; .
\end{equation}\\

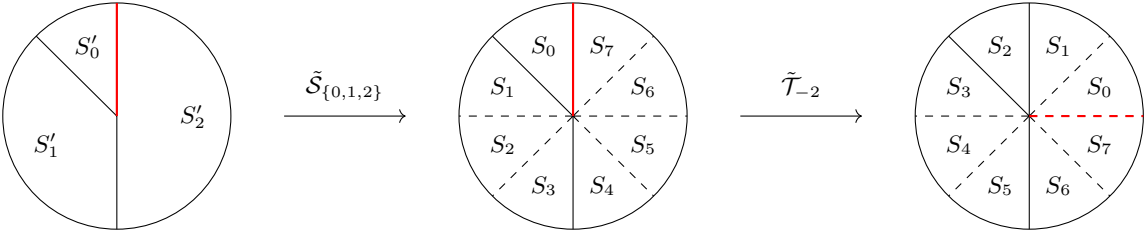
\begin{figure}[h]
\centering
 \begin{tikzpicture}
\def\R{1.5cm}
\draw (0,0) circle[radius=\R];
\draw  (0,0) --++ (90:\R)   (0,0) --++ (135:\R)      (0,0) --++ (270:\R) ;
\draw[red, thick] (0,0) --++ (90:\R);
\path ([shift=(0:1cm)]0,0)node[font=\footnotesize]{$S'_2$};
\path ([shift=(112.5:1cm)]0,0)node[font=\footnotesize]{$S'_0$};
\path ([shift=(202.5:1cm)]0,0)node[font=\footnotesize]{$S'_1$};
\draw[->] (2.2,0) -- (3.8,0);
\node[font=\footnotesize] at (3,0.4) {$\tilde{\mathcal{S}}_{\{0,1,2\}}$};
\draw (6,0) circle[radius=\R];
\draw  (6,0) --++ (90:\R)   (6,0) --++ (135:\R)      (6,0) --++ (270:\R) ;
\begin{scope}[dashed];
\draw (6,0) --++ (0:\R)  (6,0) --++ (45:\R)   (6,0) --++ (180:\R)     (6,0) --++ (225:\R)   (6,0) --++ (315:\R);
\end{scope}[dashed];
\draw[red, thick] (6,0) --++ (90:\R);
\path ([shift=(22.5:1cm)]6,0)node[font=\footnotesize]{$S_6$};
\path ([shift=(67.5:1cm)]6,0)node[font=\footnotesize]{$S_7$};
\path ([shift=(112.5:1cm)]6,0)node[font=\footnotesize]{$S_0$};
\path ([shift=(157.5:1cm)]6,0)node[font=\footnotesize]{$S_1$};
\path ([shift=(202.5:1cm)]6,0)node[font=\footnotesize]{$S_2$};
\path ([shift=(247.5:1cm)]6,0)node[font=\footnotesize]{$S_3$};
\path ([shift=(292.5:1cm)]6,0)node[font=\footnotesize]{$S_4$};
\path ([shift=(337.5:1cm)]6,0)node[font=\footnotesize]{$S_5$};
\draw[->] (8.2,0) -- (9.8,0);
\node[font=\footnotesize] at (9,0.4) {$\tilde{\mathcal{T}}_{-2}$};
\draw (12,0) circle[radius=\R];
\draw  (12,0) --++ (90:\R)   (12,0) --++ (135:\R)      (12,0) --++ (270:\R) ;
\begin{scope}[dashed];
\draw (12,0) --++ (0:\R)  (12,0) --++ (45:\R)   (12,0) --++ (180:\R)     (12,0) --++ (225:\R)   (12,0) --++ (315:\R);
\draw[red, thick] (12,0) --++ (0:\R);
\end{scope}[dashed];
\path ([shift=(22.5:1cm)]12,0)node[font=\footnotesize]{$S_0$};
\path ([shift=(67.5:1cm)]12,0)node[font=\footnotesize]{$S_1$};
\path ([shift=(112.5:1cm)]12,0)node[font=\footnotesize]{$S_2$};
\path ([shift=(157.5:1cm)]12,0)node[font=\footnotesize]{$S_3$};
\path ([shift=(202.5:1cm)]12,0)node[font=\footnotesize]{$S_4$};
\path ([shift=(247.5:1cm)]12,0)node[font=\footnotesize]{$S_5$};
\path ([shift=(292.5:1cm)]12,0)node[font=\footnotesize]{$S_6$};
\path ([shift=(337.5:1cm)]12,0)node[font=\footnotesize]{$S_7$};
\end{tikzpicture}
\caption{Pictorial representation of the action of $(\mathcal{R}^{(2, (1,3,4))})^\dagger$ on boundary conditions. The right adjoint maps $\mathcal{B}_{IR}^{(1,0)} \mapsto \mathcal{B}_{UV}^{(1,2)}$, $\mathcal{B}_{IR}^{(1,1)} \mapsto \mathcal{B}_{UV}^{(3,3)}$ and $\mathcal{B}_{IR}^{(1,2)} \mapsto \mathcal{B}_{UV}^{(4,6)}$.}\label{ExamplePictorialRepresRdagger}
\end{figure}

\noindent
To find the action of the left adjoint defined by \eqref{MFLeftAdjointRG} on boundary conditions of the IR model, we replace $P:={}^\dagger (\mathcal{R}^{(m,n)} (Y,X))$ in the above calculation. The relations on the basis vectors $e^0_a$ are again given by \eqref{Rdaggeronbcrel5}. However, the $\mathbb{Z}_d \times \mathbb{Z}_{d'}$-action on $e^0_a$ reads
\begin{equation}\label{ZZdegreesdaggerRonbc}
[e^0_a] = \big[m + \sum\nolimits_{i=1}^{a} n_i , - a + M_{IR} \big] \; .
\end{equation}
Hence, the $\mathbb{Z}_{d'}$-invariant basis vector is the one with $a = M_{IR}$, which we denote by $f := e^0_{M_{IR}}$ again.
It is of $\mathbb{Z}_d$-degree
\begin{equation}\label{ZdegreedaggerRonbc}
[f] = \big[m + \sum\nolimits_{i=1}^{M_{IR}} n_i \big] \; ,
\end{equation}
and subject to the relation
\begin{equation}
X^{\sum_{i=0}^{N_{IR}-1} n_{M_{IR} + N_{IR} -i}} f = 0 \; .
\end{equation}
Hence, the fusion product reads
\begin{equation}
{}^\dagger(\mathcal{R}^{(m,n)}) \ast_{orb} \mathcal{B}^{( N_{IR}, M_{IR})} = \mathcal{B}^{(N_{UV}, M_{UV})} \; ,
\end{equation}
with the resulting boundary condition in the UV model determined by 
\begin{equation}
M_{UV} = m + \sum\nolimits_{i=1}^{M_{IR}} n_i \quad , \quad N_{UV} = \sum\nolimits_{i=0}^{N_{IR}-1} n_{M_{IR} + N_{IR} -i} \; .
\end{equation}
In particular, the action of the adjoints on boundary conditions $\mathcal{B}^{(1, M_{IR})}$ is given by
\begin{equation}\label{actionadjointsonsinglewedges}
\begin{split}
&(\mathcal{R}^{(m,n)})^\dagger \ast_{orb} \mathcal{B}^{(1, M_{IR})} = \mathcal{B}^{(n_{M_{IR}} \; , \; m+1+\sum_{i=1}^{M_{IR}-1} n_i)} \; ,\\
&{}^\dagger(\mathcal{R}^{(m,n)}) \ast_{orb} \mathcal{B}^{(1, M_{IR})}  = \mathcal{B}^{(n_{M_{IR}+1} \; , \; m +\sum_{i=1}^{M_{IR}} n_i)} \; .
\end{split}
\end{equation}
The action of the adjoints on boundary conditions of the IR model can be represented pictorially as well. The starting point is a disk subdivided into $d'$ wedges, each wedge representing a boundary condition with $N_{IR} = 1$ in the model $X^{d'}/\mathbb{Z}_{d'}$. Again, we mark one of the lines bounding the wedges and denote the sectors by $S'_0,...,S'_{d'-1}$, starting from the marked line and going in counterclockwise direction. The boundary condition $\mathcal{B}^{(1,M_{IR})}$ then corresponds to the wedge $S'_{M_{IR}}$.
We define an operation $\tilde{\mathcal{S}}_{\{0,...,d'-1\}}$. In case we want to depict the action of $\mathcal{R}^\dagger$, this operation is defined to split each wedge $S_{M_{IR}}$ into $n_{M_{IR}}$ wedges. To depict the action of ${}^\dagger \mathcal{R}$,  $\tilde{\mathcal{S}}_{\{0,...,d'-1\}}$ is defined to split each wedge $S_{M_{IR}}$ into $n_{M_{IR}+1}$ wedges.
We define the operation $\tilde{\mathcal{T}}_{-\alpha}$ as shifting the marked line by $\alpha$ steps in clockwise direction. Then, the action of the adjoints of RG defects $\mathcal{R}^{(m,n)}$ on boundary conditions of the IR model can be represented by the operation
\begin{equation}\label{pictorialoperationrdagger}
\tilde{O}^{(m,n)} = \tilde{\mathcal{T}}_{-m} \tilde{\mathcal{S}}_{\{0,...,d'-1\}} \; .
\end{equation}
As $\sum_i n_i = d$, this operation yields a disk subdivided into $d$ wedges, see figure \ref{ExamplePictorialRepresRdagger} for an example. As before, boundary conditions with $N>1$ correspond to unions of $N$ consecutive sectors. We note that the action of RG interfaces and their adjoints on boundary conditions is consistent with \eqref{RRdagger=id} and \eqref{RdaggerR=P}.
\\

 \subsection{Commuting symmetry defects}\label{chap:commsymm}

In this section, we argue that the intertwining property \eqref{intertwining} can only be satisfied by symmetry defects and we propose a condition on the representation labels which has to be satisfied in order for symmetry defects to have this property. To do so, we make use of the pictorial representation for the action of topological defects as well as RG defects and their adjoints on boundary conditions that we described above.

\noindent
A topological defect $\mathcal{T}^{(\mathcal{N}, \mathcal{M})}$ acts on branes $\mathcal{B}^{(1,M)}$ as
\begin{equation}\label{FusionTopN1brane}
\mathcal{T}^{(\mathcal{N}, \mathcal{M})} \ast_{orb} \mathcal{B}^{(1,M)} = \mathcal{B}^{(\mathcal{N}, \mathcal{M} + M)} \; .
\end{equation}
For the intertwining property to be satisfied, both sides of equation \eqref{intertwining} have to have the same action on all branes of the UV theory. 
As we are only interested in massless theories, we consider flows from $X^d / \mathbb{Z}_d$ to $X^{d'} / \mathbb{Z}_{d'}$ with $d>d' \ge 3$. Hence, at least one wedge is annihilated by the RG defect $\mathcal{R}$ and at least three wedges survive the flow. Then, for any $\mathcal{M}$, a UV brane $\mathcal{B}^{(1,M)}$ can be found which is annihilated by $\mathcal{R}$, whereas the brane $\mathcal{B}^{(1, \mathcal{M} + M + 1)}$ or the brane $\mathcal{B}^{(1, \mathcal{M} + M)}$ is not annihilated by $\mathcal{R}$. The action of $\mathcal{T}_{IR} \ast_{orb} \mathcal{R}$ on $\mathcal{B}^{(1,M)}$ is then trivial for any $\mathcal{T}_{IR}$. On the other hand, the fusion product \eqref{FusionTopN1brane} is depicted by a union of $\mathcal{N}$ consecutive segments, starting from the segment $S_{\mathcal{M} + M}$, which means that for $\mathcal{N}>1$ the wedges $\mathcal{M} + M$ and $\mathcal{M} + M + 1$ are always contained and the action of $\mathcal{R} \ast_{orb} \mathcal{T}_{UV}^{(\mathcal{N}, \mathcal{M})}$ on $\mathcal{B}^{(1,M)}$ is non-trivial for $\mathcal{N}>1$. Hence, for $\mathcal{N}>1$, we can always find a brane for which the action of the two sides of \eqref{intertwining} differs and the intertwining property cannot be satisfied. This means that only symmetry defects can satisfy the intertwining property. Note that for the argument it was essential that some supersymmetric vacua decouple from the superconformal sector. \\
We now turn to the systematics of the preserved symmetries. On the level of matrix factorizations, we can compare the fusion products of symmetry defects with RG defects to derive a condition which has to be satisfied in order for \eqref{intertwining} to hold. The action of symmetry defects on other objects of the theory is to shift the representation label of the orbifold group. The fusion product $\mathcal{T}^{(1, \mathcal{M}_{IR})} \ast_{orb} \mathcal{R}^{(m,n)}$ then takes the form of an RG defect \eqref{RGMF} with $\mathbb{Z}_{d'}$-degrees shifted by $\mathcal{M}_{IR}$, \textit{i.e.} the generators $(f_b)_{b \in \mathbb{Z}_{d'}}$ of the module $P_0$ carry $\mathbb{Z}_{d'} \times \mathbb{Z}_d$-degree $[f_b] = [\mathcal{M}_{IR} + b, - m - \sum_{i=1}^{b} n_i]$. Likewise, the fusion product $\mathcal{R}^{(m,n)} \ast_{orb} \mathcal{T}^{(1, \mathcal{M}_{UV})}$ takes the form of an RG defect with $\mathbb{Z}_d$-charges shifted by $\mathcal{M}_{UV}$, \textit{i.e.} the basis vectors $(f_a)_{a \in \mathbb{Z}_{d'}}$ of the module $P_0$ are of $\mathbb{Z}_{d'} \times \mathbb{Z}_d$-degree $[f_a] = [a, - m - \sum_{i=1}^{a} n_i + \mathcal{M}_{UV}]$. Comparison of the fusion products then yields
\begin{equation}\label{SymmIntertwining}
\mathcal{T}^{(1, \mathcal{M}_{IR})} \ast_{orb} \mathcal{R}^{(m,n)} \simeq \mathcal{R}^{(m,n)} \ast_{orb} \mathcal{T}^{(1, \mathcal{M}_{UV})} \; ,
\end{equation}
iff the representation labels satisfy the symmetric condition
\begin{equation}\label{Symm}
\mathcal{M}_{UV} = \sum\limits_{i = b+1}^{\mathcal{M}_{IR} + b} n_i \;\;\;\;\; \forall b \in \mathbb{Z}_{d'} \; .
\end{equation}
For generic RG defects, the only solution to \eqref{Symm} is $\mathcal{M}_{UV} = 0 = \mathcal{M}_{IR}$. As expected, the identity defect $\mathcal{I}_{UV}$ remains topological and flows to the identity defect $\mathcal{I}_{IR}$ of the infrared model under generic perturbations.
In case the symmetry groups $\mathbb{Z}_d$ and $\mathbb{Z}_{d'}$ of the UV and IR models share a common subgroup $\mathbb{Z}_q$, it is possible to consider RG flows which preserve the common symmetry and non-trivial solutions to \eqref{Symm} can be found. 
RG defects which preserve a common symmetry $\mathbb{Z}_q$ can be constructed as follows.
Suppose $d$ and $d'$ have a common divisor $q \in \mathbb{N}$, with $q>1$,
\begin{equation}\label{defcommonsymm}
\frac{d}{d'} = \frac{q \tilde{d}}{q \tilde{d}'} \; . 
\end{equation}
Then, the $\mathbb{Z}_q$-symmetry can be preserved by an RG defect $\mathcal{R}^{(m,n)}$ with
\begin{equation}
n_i = n_{i + \tilde{d}'} \; \;\;\; \forall i \in \mathbb{Z}_{d'}\; .
\end{equation}
This means that the $d'$-tuple $n$ splits into $q$ equal blocks, each of length $\tilde{d}'$. As the $n_i$ have to sum up to $d$, the $n_i$ in each of the $q$ blocks have to sum up to $\tilde{d}$, \textit{i.e.}
\begin{equation}
\sum_{j=0}^{\tilde{d}'-1} n_{i+j} = \tilde{d} \quad \forall i \in \mathbb{Z}_{d'} \; .
\end{equation}
Hence, an RG defect which preserves a symmetry $\mathbb{Z}_q$ can be constructed by choosing a combination of integers $n_i \ge 1$ which satisfy
\begin{equation}
n_0 + n_1 + ... + n_{\tilde{d}' - 1} = \tilde{d} \; ,
\end{equation}
and take $q$ copies of this combination to construct the $d'$-tuple $n$. As $\tilde{d}' < \tilde{d}$, such a combination can always be found and the symmetric condition \eqref{Symm} is satisfied for all pairs
\begin{equation}
(\mathcal{M}_{UV}, \mathcal{M}_{IR}) = (\alpha \tilde{d}, \alpha \tilde{d}') \; , \;\;\; \alpha \in \mathbb{Z}_q \; .
\end{equation}
Hence, for RG flows which preserve a $\mathbb{Z}_q$-symmetry, we have $q-1$ non-trivial symmetry defects which satisfy the intertwining property in addition to the trivial solution for $\alpha = q$. 
In particular, if $d$ is a multiple of $d'$, we have $\tilde{d}' = 1$ and the full $\mathbb{Z}_{d'}$-symmetry can be preserved by flows with $n = (\tilde{d},...,\tilde{d})$.

\subsection{Generic perturbations and the weaker condition}\label{sec:generic}
Generic RG defects $\mathcal{R}^{(m,n)}$ are transparent only to the identity defect. However, explicit calculation of the fusion product \eqref{TopFlowtoIR}, which we will present below, implies that there is another symmetry defect which does not satisfy the intertwining property, yet survives the flow to the IR as well:
\begin{equation}\label{genericTopdefunderrightadjoint}
\mathcal{R}^{(m,n)} \ast_{orb} \mathcal{T}_{UV}^{(1, d-1)} \ast_{orb} (\mathcal{R}^{(m,n)})^\dagger \simeq \mathcal{T}_{IR}^{(1, d'-1)} \; .
\end{equation}
Taking the left adjoint of equation \eqref{genericTopdefunderrightadjoint} implies 
\begin{equation}\label{genericTopdefunderleftadjoint}
\mathcal{R}^{(m,n)} \ast_{orb} \mathcal{T}_{UV}^{(1,1)} \ast_{orb} {}^\dagger (\mathcal{R}^{(m,n)}) \simeq \mathcal{T}_{IR}^{(1,1)} \; .
\end{equation}
We can now construct a topological higher-${\mathcal N}$ defect by using the cone construction. Applying \eqref{boundstatestopmfs}, it is possible to form a bound state of the identity defect and the generic surviving symmetry defect which is topological again. As fusion commutes with the cone construction, we have
\begin{equation}\label{genericboundTopdefunderrightadjoint}
\mathcal{R}^{(m,n)} \ast_{orb} \mathcal{T}_{UV}^{(2, d-1)} \ast_{orb} (\mathcal{R}^{(m,n)})^\dagger \simeq \mathcal{T}_{IR}^{(2, d'-1)} \; ,
\end{equation}
and 
\begin{equation}\label{genericboundTopdefunderleftadjoint}
\mathcal{R}^{(m,n)} \ast_{orb} \mathcal{T}_{UV}^{(2,0)} \ast_{orb} {}^\dagger (\mathcal{R}^{(m,n)}) \simeq \mathcal{T}_{IR}^{(2,0)} \; .
\end{equation}
With the identification \eqref{identificationTopsCFTMF} the Landau-Ginzburg framework implies that in the corresponding B-type minimal model orbifold $\mathcal{M}_{d-2}/\mathbb{Z}_d$, in addition to the identity defect, the topological defects $\mathcal{D}_{LM}$ with $(L,M) = (0, \mp 2)$ and $(L,M) = (1, \mp 1)$ survive generic perturbations and flow to the corresponding defects in the IR,
\begin{equation}\label{ResultGenericBtwisted}
\mathcal{D}_{0, \mp 2} \rightarrow \mathcal{D}'_{0, \mp 2} \; , \;\;\; \mathcal{D}_{1, \mp 1} \rightarrow \mathcal{D}'_{1, \mp 1} \; .
\end{equation}
 The sign depends on whether the left or right adjoint of the RG defect is chosen for the mapping. The upper sign applies when choosing the right adjoint, the lower sign applies when choosing the left adjoint. We come back to a CFT interpretation of this in section \ref{sec:BackCFT}. Note that this result automatically implies that also $\mathcal{D}_{k, \pm k}$ is preserved in this sense. This follows from the conservation of   the topological defect $(0,0,2)$  by all our perturbations, as has been argued using  CFT arguments in section \ref{sec:A-picture}. If this is so, then one can use field identification to identify $(k,k,0)=(0,-2,2)=(0,-2,0) + (0,0,2)$ and conservation is implied. Note that $(0,-2,2)$ acts as spectral flow (by one unit). \\

 \noindent
We furthermore systematically verified that generically there are no other topological defects satisfying the weaker condition by implementing the defect actions on branes in a computer search. The presence of additional surviving defects is always related to symmetries.\\
\noindent
The following paragraphs give the explicit calculations verifying  \eqref{genericTopdefunderrightadjoint} and \eqref{genericTopdefunderleftadjoint}.\\

\noindent
\textbf{Composition $\mathcal{R} \ast_{orb} \mathcal{T}_{UV}^{(1, d-1)} \ast_{orb} \mathcal{R}^\dagger$}

\noindent
To verify \eqref{genericTopdefunderrightadjoint}, we consider the setup depicted in figure \ref{FigureRGflowTopdefs}. We define $Q:=(\mathcal{R}^{(m,n)}(Z,U))^\dagger$ and $P := \mathcal{R}^{(m,n)} (X,Y) \ast_{orb} \mathcal{T}_{UV}^{(1, \mathcal{M})} (Y, U)$. The matrix factorization $P$ is specified by the map $p_1^{(m,n)} (X,U)$ defined by \eqref{RGMF2} and the generators of $P_0$, $(e_a^P)_{a \in \mathbb{Z}_{d'}}$, which carry $\mathbb{Z}_{d'} \times \mathbb{Z}_d$-charge $[e_a^P] = [a, -m-\sum_{i=1}^{a} n_i + \mathcal{M}_{UV}]$. The matrix factorization $Q$ is specified by the map $(p_1^{(m,n)}(Z,U))^T$ defined by \eqref{MapAdjointsRG} and the generators of $Q_0$, $(e_b^Q)_{b \in \mathbb{Z}_{d'}}$, which are of $\mathbb{Z}_d \times \mathbb{Z}_{d'}$-degree $[e_b^Q] = [m+1+\sum_{i=1}^{b} n_i, -1-b]$. The module $P_0 \otimes Q_0$ is generated by $e_{a,b}:= e_a^P \otimes e_b^Q$. Considered as a $\mathbb{C}[X,Z]$-module, it is generated by $e^j_{a,b} := U^j e_{a,b} $ with $j \in \mathbb{N}_0$ and the basis vectors $e_{a,b}^j$ carry $\mathbb{Z}_{d'} \times \mathbb{Z}_d \times \mathbb{Z}_{d'}$-charges
\begin{equation}\label{ZZZdegreesRTRdagger}
[e^j_{a,b}] = \big[a, -m-\sum\nolimits_{i=1}^{a} n_i + \mathcal{M}_{UV} + m + 1 + \sum\nolimits_{i=1}^{b} n_i + j, -1-b \big] \; .
\end{equation}
The relations coming from the module \eqref{moduleM} read
\begin{equation}\label{R*RdaggerRel1}
X e^j_{a,b} = e^{j + n_{a+1}}_{a+1, b} \;\; \rightarrow \;\; e^{j+n_a}_{a,b} = X e^j_{a-1, b} \; ,
\end{equation}
and
\begin{equation}\label{R*RdaggerRel2}
Z e^j_{a,b} = e^{j+ n_b}_{a, b-1} \;\;\;\;\;\; \rightarrow \;\; e^{j+ n_{b+1}}_{a,b} = Z e^j_{a, b+1} \; .
\end{equation}
These relations allow to reduce the set of basis vectors to those $e^j_{a,b}$ with $0 \le j < \text{min} (n_a, n_{b+1})$. 
Setting $\mathcal{M}_{UV} = d-1$ in \eqref{ZZZdegreesRTRdagger}, the $\mathbb{Z}_d$-invariance condition yields
$j = \sum\nolimits_{i=1}^{a} n_i - \sum\nolimits_{i=1}^{b} n_i$, which can only be satisfied for $e^0_{a,a} =: f_a$. 
The basis vectors $f_a$ are of  $\mathbb{Z}_{d'} \times \mathbb{Z}_{d'}$-degree
\begin{equation}
[f_a] = [a, -a-1] \; .
\end{equation}
Combining the relations \eqref{R*RdaggerRel1} and \eqref{R*RdaggerRel2} with $a \rightarrow a+1$, the $f_a$ are subject to the relations
\begin{equation}\label{RelGenericRTRdagger}
X f_a = Z f_{a+1} \; .
\end{equation}
By comparison to \eqref{TopMF1} and \eqref{TopMF}, the $\mathbb{Z}_d$-invariant part $M^{\mathbb{Z}_d}$ of the module \eqref{moduleM} is isomorphic to the module coker$(p_1^{(1, d' - 1)})$. Hence, the resulting defect line in the IR model is isomorphic to the identity defect with left $\mathbb{Z}_{d'}$-charge shifted by $-1$, \textit{i.e.} $\mathcal{T}_{IR}^{(1,  d' - 1)}$.\\

\noindent
\textbf{Composition $\mathcal{R} \ast_{orb} \mathcal{T}_{UV}^{(1,1)} \ast_{orb} {}^\dagger \mathcal{R}$}

\noindent
To verify \eqref{genericTopdefunderleftadjoint}, we replace $Q:={}^\dagger(\mathcal{R}^{(m,n)}(Z,U))$. By \eqref{MFLeftAdjointRG}, the basis vectors $(e_b^Q)_{b \in \mathbb{Z}_{d'}}$ then carry $\mathbb{Z}_d \times \mathbb{Z}_{d'}$-charges $[e_b^Q] = [m + \sum_{i=1}^{b} n_i, -b]$, and the basis vectors $e^j_{a,b}$ are of $\mathbb{Z}_{d'} \times \mathbb{Z}_d \times \mathbb{Z}_{d'}$-degree
\begin{equation}
[e^j_{a,b}] = \big[ a, -m-\sum\nolimits_{i=1}^{a} n_i + \mathcal{M}_{UV} + m + \sum\nolimits_{i=1}^{b} n_i + j, -b \big] \; .
\end{equation}
For $\mathcal{M}_{UV} = 1$, the $\mathbb{Z}_d$-invariance condition reads $j = \sum_{i=1}^{a} n_i - \sum_{i=1}^{b} n_i -1$, which can be satisfied for $e^{n_a - 1}_{a, a-1} =: f_a$. The basis vectors $f_a$ are again subject to the relations \eqref{RelGenericRTRdagger}, however, they carry $\mathbb{Z}_{d'} \times \mathbb{Z}_{d'}$-charges
\begin{equation}
[f_a] = [a, - a + 1] \; .
\end{equation}
Comparison to \eqref{TopMF1} and \eqref{TopMF} then yields that the fusion product is isomorphic to $\mathcal{T}_{IR}^{(1,1)}$.

\subsection{Symmetric perturbations}\label{sec:symmflow}
In this section, we combine the analysis of the generic perturbation with a possible presence of additional symmetry defects. Provided the flow preserves a common symmetry $\mathbb{Z}_q$, we have
\begin{equation}\label{intertwiningsymmetriessymmflow}
\mathcal{R} \ast_{orb} \mathcal{T}_{UV}^{(1, \alpha \tilde{d})} \simeq \mathcal{T}_{IR}^{(1, \alpha \tilde{d}')} \ast_{orb} \mathcal{R} \; ,
\end{equation}
with $\tilde{d}$ and $\tilde{d}'$ defined by \eqref{defcommonsymm} and $\alpha \in \{0,...,q-1\}$. Then  the generic pairs \eqref{genericTopdefunderrightadjoint} and \eqref{genericTopdefunderleftadjoint} as well as the bound states \eqref{genericboundTopdefunderrightadjoint} and \eqref{genericboundTopdefunderleftadjoint} form orbits of length $q$ under the symmetry defects which satisfy the intertwining property. The action of a symmetry defect $\mathcal{T}^{(1, \mathcal{M})}$ on any other object of the theory is to shift the orbifold representation label by $\mathcal{M}$. Hence, in addition to \eqref{intertwiningsymmetriessymmflow}, we have
\begin{equation}\label{orbitsforrightadjoint}
\begin{split}
&\mathcal{R} \ast_{orb} \mathcal{T}_{UV}^{(1, \alpha \tilde{d} - 1)} \ast_{orb} \mathcal{R}^\dagger \simeq \mathcal{T}_{IR}^{(1, \alpha \tilde{d}' - 1)} \; ,\\
&\mathcal{R} \ast_{orb} \mathcal{T}_{UV}^{(2, \alpha \tilde{d} - 1)} \ast_{orb} \mathcal{R}^\dagger \simeq \mathcal{T}_{IR}^{(2, \alpha \tilde{d}' - 1)} \; ,
\end{split}
\end{equation}
and 
\begin{equation}\label{orbitsforleftadjoint}
\begin{split}
&\mathcal{R} \ast_{orb} \mathcal{T}_{UV}^{(1, \alpha \tilde{d} + 1)} \ast_{orb} {}^\dagger \mathcal{R} \simeq \mathcal{T}_{IR}^{(1, \alpha \tilde{d}' + 1)} \; ,\\
&\mathcal{R} \ast_{orb} \mathcal{T}_{UV}^{(2, \alpha \tilde{d} )} \ast_{orb} {}^\dagger \mathcal{R} \simeq \mathcal{T}_{IR}^{(2, \alpha \tilde{d}')} \; .
\end{split}
\end{equation}
We note that the case of generic perturbations is included and corresponds to setting $q=1$, \textit{i.e.} $\alpha = 0$. Moreover, noting that the set $-\alpha \tilde{d} = (q - \alpha)\tilde{d}$ is equivalent to the set $\alpha \tilde{d}$, the equations \eqref{orbitsforleftadjoint} can be obtained by taking the left adjoint of the equations \eqref{orbitsforrightadjoint}. Using \eqref{identificationTopsCFTMF} to translate the above results to the B-type minimal model orbifold implies that the defects with $(L,M) = (0, 2\alpha \tilde{d})$ satisfy the intertwining property and flow to the corresponding defects in the IR,
\begin{equation}\label{Result1SymmetricBtwisted}
\mathcal{D}_{0, 2\alpha \tilde{d}} \rightarrow \mathcal{D}'_{0, 2\alpha \tilde{d}'} \; .
\end{equation}
In addition, we have 
\begin{equation}\label{Result2SymmetricBtwisted}
\mathcal{D}_{0, \pm 2 \alpha \tilde{d} \mp 2} \rightarrow \mathcal{D}'_{0, \pm 2 \alpha \tilde{d}' \mp 2} \; , \;\;\; \mathcal{D}_{1, \pm 2 \alpha \tilde{d} \mp 1} \rightarrow \mathcal{D}'_{1, \pm 2 \alpha \tilde{d}' \mp 1} \; ,
\end{equation}
where the upper and lower sign again applies when choosing the right and left adjoint, respectively, for mapping the defects to the IR. 


\subsection{The maximally symmetric RG flow}
Finally, we want to consider maximally symmetric RG flows which preserve the full $\mathbb{Z}_{d'}$-symmetry of the IR model. These flows are possible for the case $d = w d'$ with $w \in \mathbb{N}$ and they are described by an RG defect $\mathcal{R}^{(m,n)}$ with $n_i = w \; \forall i \in \mathbb{Z}_{d'}$. Our calculations suggest that maximally symmetric flows preserve the complete set of UV symmetry defects.\\ 


\noindent
\textbf{Composition $\mathcal{R} \ast_{orb} \mathcal{T}^{(1, \mathcal{M}_{UV})} \ast_{orb} \mathcal{R}^\dagger$}

\noindent
We consider the setup depicted in figure \ref{FigureRGflowTopdefs} and define $P:=\mathcal{R}^{(m,n)} \ast_{orb} \mathcal{T}^{(1, \mathcal{M}_{UV})}$ and $Q:= \mathcal{R}^{(m,n) \dagger}$. For $n=(w,w,...,w)$, the generators of $P_0$, $(e_a^P)_{a \in \mathbb{Z}_{d'}}$, carry $\mathbb{Z}_{d'} \times \mathbb{Z}_d$-charges $[e_a^P] = [a, -m - aw + \mathcal{M}_{UV}]$ and the generators of $Q_0$, $(e_b^Q)_{b \in \mathbb{Z}_{d'}}$, carry $\mathbb{Z}_d \times \mathbb{Z}_{d'}$-charges $[e_b^Q] = [m+1+bw, -1-b]$. Again, we define $e_{a,b} := e_a^P \otimes e^Q_b$ and $e^j_{a,b} := U^j e_{a,b}$ with $j \in \mathbb{N}_0$. The basis vectors $e^j_{a,b}$ then carry $\mathbb{Z}_{d'} \times \mathbb{Z}_d \times \mathbb{Z}_{d'}$-charges
\begin{equation}
[e^j_{a,b}] = [a, -m - aw + \mathcal{M}_{UV} + m + 1 + bw + j, -1-b] \; .
\end{equation}
The relations coming from the module \eqref{moduleM} read
\begin{equation}
e^{j+w}_{a,b} = X e^j_{a-1, b} \quad , \quad e^{j+w}_{a,b} = Z e^j_{a, b+1} \; ,
\end{equation}
and allow to reduce the set of basis vectors to those $e^j_{a,b}$ with $0 \le j < w$. Combining the two relations, we obtain
\begin{equation}
X e^j_{a,b} = Z e^j_{a+1, b+1} \; .
\end{equation}
$\mathbb{Z}_d$-invariance singles out those basis vectors which satisfy $j = w(a-b) - \mathcal{M}_{UV} - 1$. These are given by
\begin{equation}
f_b := e^{w\alpha - \mathcal{M}_{UV} - 1}_{b+\alpha, b} \; ,
\end{equation}
with $\alpha \in \mathbb{Z}_{d'}$ and 
\begin{equation}\label{MUVdqd'}
\mathcal{M}_{UV} \in \{w\alpha - 1, w\alpha - 2, ..., w\alpha - w\} \; .
\end{equation}
They carry $\mathbb{Z}_{d'} \times \mathbb{Z}_{d'}$-charges
\begin{equation}
[f_b] = [b + \alpha, - 1- b] \; ,
\end{equation} 
and are subject to 
\begin{equation}\label{endrelmaxsymm}
X f_b = Z f_{b+1} \; .
\end{equation}
By comparison to \eqref{TopMF1} and \eqref{TopMF}, the $\mathbb{Z}_d$-invariant part $M^{\mathbb{Z}_d}$ of the module $M$ is isomorphic to the module $\text{coker}\big(p_1^{(1, \mathcal{M}_{IR} = \alpha - 1)} (X,Z) \big)$, \textit{i.e.} we have 
\begin{equation}\label{R*T*Rdaggermaxsymm}
\mathcal{R}^{(m,(w,...,w))} \ast_{orb} \mathcal{T}^{(1, \mathcal{M}_{UV})} \ast_{orb} (\mathcal{R}^{(m,(w,..,w))})^\dagger \simeq \mathcal{T}^{(1, \mathcal{M}_{IR})} \; ,
\end{equation}
for $\mathcal{M}_{UV}$ given by \eqref{MUVdqd'} and $\mathcal{M}_{IR} = \alpha - 1$.
We note that the result is in agreement with our previous considerations. The common symmetry group preserved by the flow is $\mathbb{Z}_{d'}$, \textit{i.e.} we set $q = d'$, $\alpha \in \mathbb{Z}_{d'}$, $\tilde{d} = w$ and $\tilde{d}' =1$ in section \ref{sec:symmflow}. 
Comparison to \eqref{intertwiningsymmetriessymmflow} then yields that the $d'$ symmetry defects which satisfy the intertwining property arise from the submodules built on $U^{w-1}$. Comparison to \eqref{orbitsforrightadjoint} yields, that the orbit of the generic symmetry defect arises from the submodules built on $U^0$. However, for $w>2$, the maximally symmetric flow preserves $w-2$ additional symmetry defects which form orbits of length $d'$ under the symmetry defects which satisfy the intertwining property. These orbits arise from the submodules built on $U^i$ with $0<i<w-1$.
\noindent
Taking the left adjoint of \eqref{R*T*Rdaggermaxsymm}, we also derive the equivalent relation
\begin{equation}\label{R*T*daggerRmaxsymm}
\mathcal{R}^{(m,(w,...,w))} \ast_{orb} (\mathcal{T}^{(1, \mathcal{M}_{UV})})^\dagger \ast_{orb} {}^\dagger(\mathcal{R}^{(m,(w,..,w))}) \simeq (\mathcal{T}^{(1, \mathcal{M}_{IR})})^\dagger \; ,
\end{equation}
where we used that by \eqref{adjointTops} left and right adjoints of topological defects are equivalent.

\section{Connecting back to CFT}\label{sec:BackCFT}



\subsection{Defining the left and right adjoints on the level of CFT}\label{sec:DefCFTadjoint}

In the end, we want to reconnect our discussion of symmetries, flows and defects in the TFT context back to CFT. In particular, we have seen that in generic TFTs there are two different versions of adjoint that coincide in the  case where charges are quantized and the spectral flow operator is local. The difference between TFT and CFT is indeed the insertion of a spectral flow operator at infinity. In this section we discuss the notion of "adjoint" on the level of CFT defects. Here, we interpret the defect as an operator on the closed string Hilbert space, meaning as a map
\begin{equation}
D: {\cal H}^{(1)} \to {\cal H}^{(2)} \ ,
\end{equation}
where ${\cal H}^{(1)}$ and ${\cal H}^{(2)}$ are the Hilbert spaces of the two theories connected by the defect. In this section, we take the "mirror" perspective, and the Hilbert space is the Hilbert space of the minimal model with diagonal modular invariant and the defects we are interested in are A-type. As an operator on a Hilbert space, a defect $D$ has an adjoint in the usual CFT-sense that we denote by $D^*$ and it maps
\begin{equation}
D^*: {\cal H}^{(2)} \to {\cal H}^{(1)} \; .
\end{equation}
To preserve A-type supersymmetry, the defect has to intertwine the action of the supersymmetry currents of the N=2 algebra, see \cite{Bachas:2013nxa} for a discussion of this point and see \cite{Hori:2000ck_HIV} for a discussion of adjoints of boundary states that is similar to the following discussion.
For A-type gluing conditions, it has to satisfy
\begin{equation}
D\left( G^{(1)\pm } -i \tilde{G}^{(1) \mp} \right)=\left( G^{(2)\pm } -i \tilde{G}^{(2) \mp} \right) D \; .
\end{equation}
By taking the adjoint of this equation, we derive that the ordinary CFT adjoint operator $D^*$ satisfies 
\begin{equation}
\left( G^{(1)\pm } +i \tilde{G}^{(1) \mp} \right) D^* = D^* \left( G^{(2)\pm } + i \tilde{G}^{(2) \mp} \right) \; .
\end{equation}
So the CFT adjoint operator does not preserve the same supersymmetry. To compare with the topologically twisted theory, we need the same supersymmetry to be preserved by the operator, therefore look for a modified version of adjoint. Let us define the following operator:
\begin{equation}
D^\dagger = e^{\pi i \tilde{Q}^{(1)}} D^* e^{-\pi i \tilde{Q}^{(2)}} \; .
\end{equation}
Here, $\tilde{Q}^{(1)}$ denotes the operator that gives the right moving $U(1)$ charge when acting on states in the Hilbert space of theory 1, and likewise $\tilde{Q}^{(2)}$. In the minimal model, $\tilde{Q}=Q$ on states, so we do not play with the possibility to consider the left moving $U(1)$ charges. With this newly defined operator, we can verify that the adjoint satisfies the same gluing conditions as the original defect,
\begin{equation}
\left( G^{(1)\pm } -i \tilde{G}^{(1) \mp} \right) D^\dagger = D^\dagger \left( G^{(2)\pm } - i \tilde{G}^{(2) \mp} \right) \ ,
\end{equation}
as desired. We furthermore note that our definition is compatible with the composition of operators. To see this, consider two  defect-operators
\begin{equation}
D_1: {\cal H}^{(1)} \to {\cal H}^{(2)} , \quad D_2: {\cal H}^{(2)} \to {\cal H}^{(3)} \; .
\end{equation}
It is then easy to see that
\begin{equation}
(D_2 D_1)^\dagger = D_1^\dagger D_2^\dagger \; ,
\end{equation}
as a consequence of the properties of the CFT adjoint $(D_2 D_1)^*= D_1^* D_2^*$. To make further contact with the properties of adjoints in the TFT, we need a corresponding left adjoint ${}^\dagger D$ such that
\begin{equation}
{}^\dagger D^\dagger = D \; .
\end{equation}
This operation is provided by
\begin{equation}
{}^\dagger D = e^{-\pi i \tilde{Q}^{(1)}} D^* e^{+\pi i \tilde{Q}^{(2)}} \ .
\end{equation}
Similarly to the defect $D^\dagger$, also ${}^\dagger D$ satisfies the same gluing conditions as $D$. Moreover, in the case where the $U(1)$ charges are integer, the two versions of adjoint agree. This reproduces a further property of the left and right adjoints from TFT. Note  that our CFT discussion would imply that for topological defects (that preserve the full symmetry algebra) within any theory, the left and right adjoints are the same, since they commute with the charge operator. We have checked in our discussion of topological defects as equivariant matrix factorizations in section \ref{sec:MFTopDefs} that this is indeed the case for minimal models. 

\subsection{Revisiting the results from the TFT discussion}

Recall the results from the previous sections,
\begin{equation}
R \ast T \ast R^\dagger = T' \; ,
\end{equation}
where $T$ is the symmetry defect with Cardy labels $(0,-2,2)$ in the UV theory, and $T'$ in the IR theory. Taking the left adjoint, we obtain
\begin{equation}
R \ast T^\dagger \ast ({}^\dagger R) = T^{' \dagger} \ ,
\end{equation}
where we used that for topological defects left and right adjoints agree. Using the above discussion, we can now read the above equations on the operator level in CFT as well. On this level, we can identify the symmetry defect  $(0,-2,2)$ with the operator $e^{-2 \pi i \tilde{Q}} = e^{-\pi i (\tilde{Q} + Q)}$ in either theory. Plugging in all our identifications, we obtain on the CFT level the equations
\begin{equation}
R e^{\pm \pi i \tilde{Q}} R^* = e^{\pm \pi i \tilde{Q}} \; .
\end{equation}
Note that the two operators $\exp (\pm \pi i \tilde{Q})$ are related by the usual CFT adjoint. The above equation expresses that the notion of $(-1)^{\tilde{F}}$ is compatible under the perturbations we are considering and the same on the two sides, even in the non-CY case.
The discussion in section \ref{sec:DefCFTadjoint} shows that the preservation of $e^{\pm \pi i \tilde{Q}}$ that we have observed in the explicit example holds more generally, in the cases that  a topological subsector can be extracted in a consistent way.


\section{Conclusion and Outlook}\label{sec:conclusion}

In this paper we elaborated on the relation between supersymmetric conformal field theories and their twisted versions, which are topological field theories. Our main example were $N=(2,2)$ minimal models, which have been identified with Landau-Ginzburg models in \cite{Vafa:1988uu}, where no boundaries or defects were considered. The CFT-LG correspondence was later extended to the case of boundaries in \cite{Kapustin:2003rc, Brunner:2003dc} and to the case of defects in \cite{Brunner:2007qu}. 

RG flow of minimal models were described in terms of defects in \cite{Brunner:2007ur}. On the level of the topological theory, the RG defect contains the information on which elementary branes decouple in the IR, and also determines the decay of non-elementary branes along the flow. 

We combine all pieces and obtain results on flows on the CFT as well as TFT levels, in particular on the behavior of topological defects. First of all, in section~\ref{chap:commsymm} we identify symmetry defects to which the perturbation is entirely invisible; they commute with the RG defect describing the flow. Furthermore, in section~\ref{sec:generic} we identify a particular symmetry defect, related to spectral flow,  that remains invariant under all flows in all models. As a consequence there is also a topological defect corresponding to a non-linear matrix factorization, that in particular is not a symmetry defect. In case there are commuting symmetry defects, they act naturally on the preserved defects, and the preserved defects form orbits under the action of the symmetry defects.

As we mentioned, and as is discussed in more detail in \cite{Carqueville:2012st}, on the level of topological Landau-Ginzburg orbifold models, defects have a right and also a left adjoint. These two notions of adjoint are in general not isomorphic -- as opposed to adjoints of operators in conformal field theory. This indicates that the difference is related to the insertion of spectral flow, and we clarify this in section \ref{sec:BackCFT}. Our discussion implies that there are classes of LG orbifold defects for which left and right adjoints agree, namely those preserving extended symmetries, such as the  defects corresponding to those preserving the full rational symmetry in our model.\\

\noindent
\textbf{Acknowledgements:} The work of I.B. is supported by the DFG project "Defekte und nicht perturbative Operatoren in superkonformen Theorien", as well as the Excellencecluster Origins. We thank Kentaro Hori for useful discussions and Nils Carqueville for comments on the manuscript.



\bibliographystyle{JHEP}
\bibliography{references}

\end{document}